\documentclass{article}

\usepackage{arxiv}

\usepackage[utf8]{inputenc} 
\usepackage[T1]{fontenc}    
\usepackage{hyperref}       
\usepackage{url}            
\usepackage{booktabs}       
\usepackage{amsfonts}       
\usepackage{nicefrac}       
\usepackage{microtype}      
\usepackage{lipsum}		
\usepackage{graphicx}
\usepackage[numbers]{natbib}
\usepackage{doi}
\usepackage{tikz}
\usetikzlibrary{decorations.markings}
\usetikzlibrary{decorations.pathreplacing}
\usetikzlibrary{shapes.geometric}
\usepackage{mathtools}
\usepackage{amsmath,amssymb,bm}
\usepackage{amsthm}
\newtheorem*{hypothesis}{Hypothesis}
\usepackage{mathrsfs}
\usepackage{pgfplots}
\usepackage{caption}
\usepackage[linesnumbered,ruled,vlined]{algorithm2e}
\graphicspath{ {Images/} }
\usepackage{multirow}
\usepackage{makecell}
\usepackage{float}
\usepackage[multiple]{footmisc}
\usepackage{hyperref}
\hypersetup{
    colorlinks=true,
    linkcolor=blue,
    filecolor=magenta,
    urlcolor=cyan,
}

\title{Gradient-Annihilated PINNs for Solving Riemann Problems: Application to Relativistic Hydrodynamics}

\date{\footnotesize{May 3, 2023}}	

\author{ \href{https://orcid.org/0000-0002-2305-5261}{\includegraphics[scale=0.06]{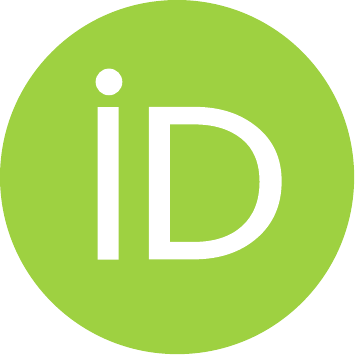}\hspace{1mm}\footnotesize{Antonio Ferrer-Sánchez}}\thanks{\scriptsize{IDAL, Electronic Engineering Department, ETSE-UV, University of Valencia, Avgda. Universitat s/n, 46100 Burjassot, Valencia, Spain.}}\textsuperscript{\textnormal{,}}\thanks{\scriptsize{Valencian Graduate School and Research Network of Artificial Intelligence (ValgrAI), Spain.}} \\
	\footnotesize{\texttt{anfesan6@uv.es}} 
	\And
	\href{https://orcid.org/0000-0001-9378-0285}{\includegraphics[scale=0.06]{orcid.pdf}\hspace{1mm}\footnotesize{José D. Martín-Guerrero}}\textsuperscript{\textnormal{1,2}}\\
	\footnotesize{\texttt{jose.d.martin@uv.es}} \\
	\And
	\href{https://orcid.org/0000-0003-3688-9609}{\includegraphics[scale=0.06]{orcid.pdf}\hspace{1mm}\footnotesize{Roberto Ruiz de Austri}} \thanks{\scriptsize{Instituto de Física Corpuscular CSIC-UV, c/Catedrático José Beltrán 2, 46980 Paterna, Valencia, Spain.}} \\
 \footnotesize{\texttt{rruiz@ific.uv.es}} \\
	\And
	\href{https://orcid.org/0000-0001-8709-5118}{\includegraphics[scale=0.06]{orcid.pdf}\hspace{1mm}\footnotesize{Alejandro Torres-Forné}} \thanks{\scriptsize{Departamento de Astronomía y Astrofísica, Universitat de Valencia, Dr. Moliner 50, 46100, Burjassot (Val\`encia), Spain.}}\textsuperscript{\textnormal{,}}\thanks{\scriptsize{Observatori Astronòmic, Universitat de València, Catedr\'atico 
  Jos\'e Beltr\'an 2, 46980, Paterna (Val\`encia), Spain.}}\\
 \footnotesize{\texttt{alejandro.torres@uv.es}} \\
	\And
	\href{https://orcid.org/0000-0001-6650-2634}{\includegraphics[scale=0.06]{orcid.pdf}\hspace{1mm}\footnotesize{Jos\'e A. Font}}\textsuperscript{\textnormal{4,5}} \\
 \footnotesize{\texttt{j.antonio.font@uv.es}} \\
}

\renewcommand{\headeright}{}
\renewcommand{\undertitle}{A Preprint}

\hypersetup{
pdftitle={Gradient-Annihilated PINNs for Solving Riemann Problems: Application to the Relativistic Hydrodynamics},
pdfkeywords={Riemann problem, Euler equations, Machine learning, Neural networks, Relativistic hydrodynamics},
}

\begin{document}
\maketitle

\begin{abstract}
We present a novel methodology based on Physics-Informed Neural Networks (PINNs) for solving systems of partial differential equations admitting discontinuous solutions. Our method, called Gradient-Annihilated PINNs (GA-PINNs), introduces a modified loss function that requires the model to partially ignore high-gradients in the physical variables, achieved by introducing a suitable weighting function. The method relies on a set of hyperparameters that control how gradients are treated in the physical loss and how the activation functions of the neural model are dynamically accounted for. The performance of our GA-PINN model is demonstrated by solving Riemann problems in special relativistic hydrodynamics, extending earlier studies with PINNs in the context of the classical Euler equations. The solutions obtained with our GA-PINN model correctly describe the propagation speeds of discontinuities and sharply capture the associated jumps. We use the relative $l^{2}$ error to compare our results with the exact solution of special relativistic Riemann problems, used as the reference ``ground truth'', and with the corresponding error obtained with a second-order, central, shock-capturing scheme. In all problems investigated, the accuracy reached by our GA-PINN model is comparable to that obtained with a shock-capturing scheme and significantly higher than that achieved by a baseline PINN algorithm. An additional benefit worth stressing is that our PINN-based approach sidesteps the costly recovery of the primitive variables from the state vector of conserved variables, a well-known drawback of grid-based solutions of the relativistic hydrodynamics equations. Due to its inherent generality and its ability to handle steep gradients, the GA-PINN methodology discussed in this paper could be a valuable tool to model relativistic flows in astrophysics and particle physics, characterized by the prevalence of discontinuous solutions.
\end{abstract}


\section{Introduction}
\setcounter{footnote}{0}

Within the rapidly-evolving field of computer science, 
Deep Learning (DL) has firmly established itself as a useful tool for solving a wide range of problems, with applications in areas as diverse as e.g.~image processing \citep{ref44,ref45}, natural language processing \citep{ref43}, machine translation \citep{ref46} and time series prediction \citep{ref47}. Technologies involved include recurrent neural networks (RNNs) \citep{ref10}, such as long-term memory architectures (LSTMs) \citep{ref6}, attentional mechanisms \citep{ref7}, encoder-decoder architectures \citep{ref8,ref9}, etc. One of the areas in which DL has shown particularly promising results is in the evolution of physical systems by solving the partial differential equations (PDEs) that describe them. PDEs form the basis for the mathematical modelling of many physical phenomena, as they determine how a system should behave and evolve. Therefore, their understanding and accurate solution are crucial to fully describe the underlying physics. With this in mind, physics-informed neural networks (PINNs) \citep{ref4} have recently emerged as a general-purpose algorithm for solving PDEs that allow to capture the behaviour of physical systems, making them a promising alternative to traditional numerical methods. 

Since their first appearance in the scientific literature \citep{ref20,ref21,ref4}, PINNs have managed to position themselves as one of the most advanced and widely used numerical PDE solvers based on machine learning \citep{ref22,ref23}. PINNs can therefore be considered as a competitive approach that provides an easy-to-use framework for solving PDEs using neural networks (NNs) as approximators. In this way, PINNs are able to compute the solutions of PDEs by incorporating physical information into both the underlying architecture and the training procedure by learning from data and from the physical laws of the problem. The versatility of PINNs in understanding and solving physical systems has led to the increasing popularity of these models in the field of machine learning research and their direct application in various scientific areas \citep{ref24,ref25}.

On the other hand, the universal approximation theorem for NNs \citep{ref26, ref27, ref28} guarantees that any continuous PDE can be approximated, and thus solved, by a NN. This is achieved primarily by the fact that the information contained in the equations is incorporated into the loss function of the NN, which is minimised during training. In general, this loss metric simultaneously contains both information about the physical equations and the initial and boundary conditions of the problem. Nevertheless, there is nowadays no theoretical background 
that 
guarantees that the ability of NNs to solve any continuous PDE can be directly generalised to discontinuous situations. Therefore, the investigation of PDEs admitting discontinuous solutions using PINNs 
requires modifications to the basic algorithms of these networks to ensure an accurate representation of the equations.

In the recent literature on solving such type of PDEs with PINNs, different and novel methods have been proposed to guide the basic PINN algorithm to capture discontinuous features. The classical compressible Euler equations constitute a recurrent example for those studies \citep{ref19}. These are equations that describe the behaviour of the flow of a fluid or gas as a nonlinear hyperbolic system of conservation laws for the mass, momenta, and energy. The development of discontinuous solutions even for smooth initial data is a consequence of the nonlinear hyperbolic nature of the equations. Discontinuities are also commonly introduced through Riemann problems (or shock tubes), the simplest initial-value-problem with discontinuous initial data. A well-known example is the \textit{Sod Shock Tube} problem \citep{ref29}. This problem was investigated using PINNs in \citep{ref24} and \citep{ref25} that proposed suitable modifications to the basic framework of the NNs to capture the jumps. In \citep{ref30} variations of the traditional PINNs method to solve other problems with discontinuous initial conditions besides the \textit{Sod Shock Tube} problem were also presented. Those modifications were based on an extension of the domain used as training data and the strategical weighting of each part of the loss function. We note that in addition to the Euler equations, there are other systems that can also develop discontinuities and have been investigated using PINNs, such as e.g.~Burgers equation and the convection-diffusion equation \citep{ref31}.


In this paper we investigate the performance of PINNs when solving the equations of relativistic hydrodynamics (RH). This system is a generalisation of the classical Euler equations of fluid dynamics and can also be cast as a nonlinear hperbolic system of conservation laws (see e.g.~\cite{Font:1994,ref32,ref39,FontLRR} and references therein). In the limit of special relativity, these equations describe  situations in which the velocity of the fluid approaches the speed of light, while the full theory of general relativity must be employed when describing systems in which the gravitational field is sufficiently strong or the self-gravity of the fluid is significant. Such extreme conditions appear in e.g.~particle physics and astrophysics (particularly in the description of compact astrophysical objects). To the best of our knowledge this paper presents the first proposal and development of a framework for a PINN-based numerical solution of the equations of special relativistic hydrodynamics. Here, we put forward a novel PINN method, which we call Gradient-Annihilated PINN (GA-PINN), which is based on the construction of a general loss function able to naturally detect discontinuities in a physical spacetime domain. The discontinuities are detected up to a certain degree which is controlled by introducing a factor that directly depends on the gradient of the particular fluid variable. Furthermore, the NN architecture proposed in this work incorporates dynamic activation functions that are parameterised to facilitate learning through minimisation procedures while accounting for discontinuities. Our GA-PINN method is assessed with a number of Riemann problems with analytic solution and, in addition, its performance is compared against the results provided by a second-order central high-resolution shock capturing scheme~\cite{ref31}. As we show below our method yields competitive results for solving the equations of relativistic hydrodynamics in the presence of discontinuities, accurately capturing the correct propagation speeds and the associated jumps in the fluid variables.



The rest of the paper is organized as follows: Section~\ref{sec:back} describes the background of our research with a literature review of PINNs, relativistic hydrodynamics, and the application of PINNs to classical hydrodynamics. Section~\ref{sec:methods} outlines the methodology of our GA-PINN  algorithm. The numerical experiments used to assess our method and the corresponding results are presented in   Section~\ref{sec:results}. Finally, in Section~\ref{sec:discuss} we discuss the  main conclusions of this work and mention possible suggestions for future work. Code for training examples has been provided by the authors\footnote{\href{https://github.com/Netzuel/GA_PINNs_Repository}{GA-PINNs GitHub repository.}}.

\section{Background}
\label{sec:back}
\subsection{Physics-Informed Neural Networks}
\label{sec:PN}

The basic methodology of PINNs \citep{ref4} uses neural networks to solve a system of PDEs. In order to achieve this objective, the concept of inductive biases is employed, which encompasses the implicit assumptions that the neural model incorporates concerning the underlying physics of the problem under consideration. By encoding such information via the network architecture, loss function, initial conditions, and other pertinent physical properties, it is possible to achieve model convergence and, in certain instances, enhance their performance. The algorithm uses the automatic differentiation available in current frameworks \citep{ref35} to construct the physical derivatives of the variables of interest with respect to the domain coordinates. In this way, it is possible to evaluate the PDEs system and perform a process of minimization and training of the architecture parameters. In general, one can consider a system of PDEs for a set of variables $\mathcal{U}:=\mathcal{U}(t,\bm{x})$ parameterized by the set of parameters $\Lambda$, on a certain domain of interest $\Omega$ and a certain time interval $[0,T]$:
\begin{equation}
\mathcal{F}\left(t,\bm{x};\mathcal{U};\frac{\partial\mathcal{U}}{\partial t},\frac{\partial^{2}\mathcal{U}}{\partial t^{2}},...;\frac{\partial\mathcal{U}}{\partial x_{1}},...,\frac{\partial\mathcal{U}}{\partial x_{D}};\frac{\partial\mathcal{U}}{\partial x_{1}\partial x_{1}},...,\frac{\partial\mathcal{U}}{\partial x_{1}x_{D}};...;\Lambda\right)=0,\qquad\bm{x}=(x_{1},...,x_{D})\in\Omega,\qquad t\in[0,T],
\label{eq1}
\end{equation}
in conjunction with a set of boundary conditions
\begin{displaymath}
\mathcal{B}(t,\bm{x})=0\quad\text{with}\quad(t,\bm{x})\in(0,T]\times\partial\Omega,
\end{displaymath}
and, in general, a set of initial conditions
\begin{displaymath}
\mathcal{IC}(t,\bm{x})=0\quad\text{with}\quad(t,\bm{x})\in\{0\}\times\Omega.
\end{displaymath}
The operator $\mathcal{F}$ defined in (\ref{eq1}) can be understood as the part containing each of the $k$ physical constraints of the system, $\mathcal{R}_{k}(t,\bm{x})$. On the other hand, $\mathcal{B}$ and $\mathcal{IC}$ represent the sets of domain points that are known a priori and are conditions that will be required in the training of the underlying neural network. The architecture consists of a certain set of parameters $\Theta$, whose output will be represented as $\mathcal{U}(t,\bm{x};\Theta):=\mathcal{U}_{\Theta}(t,\bm{x})$. In this way, through the training procedure, an attempt will be made to obtain a vector of physical variables as similar as possible to the true one.
\begin{displaymath}
\mathcal{U}_{\Theta}(t,\bm{x})\approx\mathcal{U}(t,\bm{x}).
\end{displaymath}
In order to obtain the solution, a loss function is constructed as the sum of several individual terms, each of them defined on a different domain.
\begin{equation}
\mathcal{L}=\sum_{i}\omega_{i}\mathcal{L}_{i}:=\omega_{\mathcal{IC}}\mathcal{L}_{\mathcal{IC}}+\omega_{\mathcal{B}}\mathcal{L}_{\mathcal{B}}+\omega_{\mathcal{R}}\mathcal{L}_{\mathcal{R}}.
\label{eq2}
\end{equation}
Each one of the addends in (\ref{eq2}) can be defined  as a \textit{Mean Squared Error} (MSE) (or as another metric function such as MAE), as follows:
\begin{gather*}
\mathcal{L}_{\mathcal{IC}}:=\frac{1}{N_{\mathcal{IC}}}\sum_{\{0\}\times\Omega}\mathopen|\mathcal{U}_{\Theta}(0,\bm{x})-\mathcal{U}(0,\bm{x})\mathclose|^{2},\\
\mathcal{L}_{\mathcal{B}}:=\frac{1}{N_{\mathcal{B}}}\sum_{(0,T]\times\partial\Omega}\mathopen|\mathcal{U}_{\Theta}(t,x\in\partial\Omega)-\mathcal{U}(t,x\in\partial\Omega)\mathclose|^{2},\\
\mathcal{L}_{\mathcal{R}}:=\frac{1}{N_{\mathcal{R}}}\sum_{(0,T]\times\Omega}\underbrace{\sum_{k}\mathopen|\mathcal{R}_{k}(t,x;\Theta)\mathclose|^{2}}_{\mathopen|\mathcal{F}\mathclose|^{2}},
\end{gather*}
where the vector $\bm{\omega}=(\omega_{\mathcal{IC}},\omega_{\mathcal{B}},\omega_{\mathcal{R}})$ represents the weights that indicate in what proportion the terms in (\ref{eq2}) are mixed. These weights can be static numbers, vectors, or even dependent functions of the physical variables. In addition to these terms, others can also be considered, such as a loss factor related to experimental measurements of the variables, if any. The method of incorporating initial and boundary conditions by allowing the network to learn them concurrently with the differential equations, as opposed to mandating their usage from the onset, is referred to as \textit{soft enforcement} in the literature. Conversely, the approach of compelling the network to strictly adhere to such conditions from the outset is termed \textit{hard enforcement} \citep{ref36}.

\subsection{Classical and relativistic hydrodynamics}
\label{section22}
In physics, the Navier-Stokes equations are a set of partial differential equations that describe the behaviour of a fluid with non-zero viscosity. Within this framework, when considering a non-viscous fluid in which the dissipative components are negligible compared to the convective ones, the so-called Euler equations are obtained. The Euler equations represent the conservation of mass, momentum and energy of an inviscid fluid, and can be written in the following conservative form
\begin{equation}
\frac{\partial\mathcal{U}}{\partial t}+\nabla\cdot f(\mathcal{U})=\mathcal{S}(\mathcal{U}),
\label{eq3}
\end{equation}
where~$\mathcal{U}=\mathcal{U}(t,\bm{x})$ is a vector of physical variables with $\bm{x}\in\Omega\subset\mathbb{R}^{D}$ and $t\in[0,T]$, being $D$ the spatial dimensionality of the problem. Furthermore, $f(\mathcal{U})$ is the flux vector which depends on the conserved variables $\mathcal{U}$, and $\mathcal{S}(\mathcal{U})$ is a vector accounting for possible source terms. This way of writing the system helps emphasize that a certain property of the system is conserved, as is the case of hyperbolic systems such as the Euler equations \citep{ref37}. Particularizing to the state vector $\mathcal{U}=(\rho,j,e)$ the classical Euler equations are obtained, which in one spatial dimension and using Cartesian coordinates can be written as
\begin{equation}
\frac{\partial}{\partial t}
\begin{pmatrix}
\rho\\
j\\
e
\end{pmatrix}+\frac{\partial}{\partial x}
\begin{pmatrix}
j\\
p+\frac{j^{2}}{\rho}\\
\frac{j}{\rho}(e+p)
\end{pmatrix}=0,
\label{eq4}
\end{equation}
where $\rho$ is the density of the fluid, $j=\rho u$ is its momentum density (along direction $x$), $u$ refers to the velocity and $e$ corresponds to the total energy density (total energy per unit volume). Note that no source terms appear in the above system, a consequence of using Cartesian coordinates. Along with (\ref{eq4}), it is necessary one more equation in order to close the system, that is, the equation of state (EOS) describing the relationship between the pressure of the fluid and its density and energy, i.e. $p=p(\rho,e)$. One of the simplest EOS to consider is that of an ideal gas, given by 
\begin{equation}
p=(\Gamma-1)\left(e-\frac{1}{2}\frac{j^{2}}{\rho}\right),
\label{eq5}
\end{equation}
where $\Gamma$ is the adiabatic index. 

The classical Euler equations (\ref{eq4}) need to be modified when describing the dynamics of fluids that involve high energies, as those fluids require a relativistic treatment. Such situations are encountered in extreme astrophysical systems characterized by the presence of compact objects as neutron stars, black holes, micro-quasars, extragalactic jets, as well as in high-energy collisions of heavy ions (see e.g.~\citep{FontLRR,ref34} and references therein). Therefore, the equations of relativistic hydrodynamics govern the dynamics of fluids that are immersed in relativistic conditions which involve one (or both) of the following situations:
\begin{itemize}
\item Scenarios in which the fluid as a whole (or parts of it) acquire velocities close to the speed of light in vacuum ($c$), i.e. high-speed flows.
\item Situations in which there is a gravitational field that is strong enough to consider the effects of the curvature of the spacetime in the fluid motion.
\end{itemize}
The investigation of the dynamics of these astrophysical systems strongly relies on numerical simulations. The use of accurate numerical methods which exploit the conservative form of the relativistic hydrodynamics equations is essential for the modelling. Among those, the so-called high-resolution shock-capturing schemes (HRSC)~\citep{ref38} constitute a favoured class. Similar to the classical Euler equations, the general relativistic hydrodynamics equations (GRHD) also correspond to local conservation laws of the matter current density and of momentum and energy. In order to write and derive these equations, one needs to first take the components of the bilinear form (\textbf{T}), which corresponds to the energy-momentum tensor, and of the four-vector $\bm{J}$, which represents the current of rest mass,
\begin{equation}
J^{\mu}=\rho u^{\mu},\qquad T^{\mu\nu}=\rho hu^{\mu}u^{\nu}+pg^{\mu\nu},\qquad h:=1+e+\frac{p}{\rho},
\label{eq6}
\end{equation}
where $\rho$ is the rest-mass density of the fluid, $p$ the pressure, $h$ the specific enthalpy, $e$ the specific internal energy, $u^{\mu}$ the four-velocity of the fluid, and $g^{\mu\nu}$ corresponds to the metric of the Riemannian spacetime manifold, $\mathscr{M}$, in which the fluid evolves. Note that in the previous equations (and elsewhere in this paper) we are using natural units with $c=1$.

In conjunction with these expressions, it is necessary, as in the classical case, to include a thermodynamic relationship that connects the pressure ($p$), density ($\rho$) and internal energy ($e$) of the fluid in order to close the system, i.e.~the EOS. For the modelling of astrophysical systems an ideal gas EOS is often employed. In analogy with the classical case (\ref{eq5}), this EOS can be written as
\begin{equation}
p=(\Gamma-1)\rho e,
\label{eq7}
\end{equation}
where $\Gamma$ is the adiabatic index. 

For numerical applications, the covariant equations (\ref{eq6}) must be expressed with respect to a particular coordinate system. To do so, a widely used approach in general relativity is  the so-called $\{3+1\}$ formulation (also called the Cauchy formulation) which allows foliating the spacetime with a set of non-intersecting spacelike hyper-surfaces with spatial coordinates $x^i$ $(i=1,2,3)$ at constant coordinate time $x^0$. This allows to derive a conservative formulation of the GRHD equations in the 3+1 split of the spacetime in terms of a general metric tensor and the associated mathematical connection~\cite{ref32}. In this paper we focus on Minkowski spacetime in four dimensions ($\mathscr{M}^{4}$), i.e.~the flat spacetime of special relativity, and particularize the GRHD equations in this framework (see~\cite{Font:1994} for details). For the derivation of a  general formulation, the interested reader is addressed to \citep{ref32} which reports a theoretical development widely used in the literature to obtain the GRHD equations in conservation form. With this in mind, the relativistic Euler equations written in Minkowski spacetime in one spatial dimension in Cartesian coordinates (no source terms) read
\begin{equation}
\frac{\partial}{\partial t}
\begin{pmatrix}
D\\
S\\
\tau
\end{pmatrix}+\frac{\partial}{\partial x}
\begin{pmatrix}
Du\\
Su+p\\
S-Du
\end{pmatrix}=0.
\label{eq8}
\end{equation}
In this formulation, the vector of the conserved quantities is $\mathcal{U}:=(D,S,\tau)$. Its components are the relativistic densities of mass, momentum (along direction $j$) and energy, respectively, defined as
\begin{equation}
\label{eq9}
\begin{gathered}
D=\rho W,\\
S_{j}=\rho h W^{2}u_{j},\\
\tau=\rho hW^{2}-p-D,
\end{gathered}
\end{equation}
where $W$ is the Lorentz factor
\begin{displaymath}
W^{2}:=\frac{1}{1-\sum_{j=1}^{3}u_{j}^{2}},
\end{displaymath}
and where, in general, the index $j$ runs from 1 to 3 corresponding to the spatial coordinates of the representation. In addition, the specific enthalpy, $h$, can be written as $h=1+\Gamma e$ for our specific case of an ideal gas by using its definition in (\ref{eq6}) and the EOS in (\ref{eq7}).

Obtaining accurate numerical solutions of relativistic fluids in the presence of strong shock waves is fundamental to guarantee progress in the investigation of dynamical systems in high-energy physics and astrophysics (see \citep{FontLRR,ref39} and references therein). Shock waves are a fairly common feature in nonlinear hyperbolic equations, appearing even when the initial data are smooth, and the relativistic Euler equations (\ref{eq8}) are no exception. Due to the presence of the Lorentz factor, the relativistic version of the Euler equations are much more coupled than their classical limit (\ref{eq4}). Experience has shown that traditional numerical schemes benefit greatly from using their conservative form to overcome certain numerical limitations, especially when dealing with ultra-relativistic flows ($W\sim 10$ and higher). Modern HRSC schemes have long shown to be perfectly suited for that purpose \citep{FontLRR}.

A key difference between the classical (\ref{eq4}) and the relativistic (\ref{eq8}) Euler equations is the fact that in the latter there is no analytical expression that allows to write the physical or primitive variables, $(\rho,u,p)$, in terms of the conserved ones, $(D,S,\tau)$. This implies that at every time step of a time-dependent numerical simulation it is always  necessary to retrieve the primitives implicitly (in order to solve Riemann problems), which reduces the efficiency of codes based on finite differences (as is the case of HRSC schemes). This well-known drawback can be, however, sidestepped using the PINN-based methodology discussed in this work. The reason is that the networks are capable of working with both sets of variables simultaneously in the internal calculations they perform. In this paper, in order to make a direct comparison with numerical results based on HRSC schemes  available in the literature, our DL-based methodology will also employ the conservative form of the equations shown in (\ref{eq8}).
 
\subsection{Related work}
The basic PINNs algorithm described above has proven to be very useful in obtaining the solution of many physical systems using DL. Nevertheless, PINNs, especially when these problems involve the propagation of shock waves or discontinuities in the initial and/or boundary conditions, may not produce results as satisfactory as conventional numerical schemes, or may even fail to converge. On the other hand, depending on the neural architecture considered, these models may also vary their performance depending on different hyperparameters that must be chosen a priori, such as the number of hidden layers and the number of neurons in them, their distribution and interconnection, the activation functions considered, the learning rate and optimizer chosen, etc. 

In \citep{ref40} a comprehensive theoretical study was performed to understand how the gradient of the loss function with respect to the neural architecture parameter set ($\nabla_{\Theta}\mathcal{L})$ is transferred along it. The authors concluded that these gradients of the loss term, corresponding to the boundary and initial conditions, tend to vanish as the NN is traversed, assuming that it is necessary to find a balance in each of the terms of the final function. This, together with an analysis of the corresponding Hessian matrix and learning rate, led the authors to define adaptive weights, i.e. the set $\bm{\omega}$ in (\ref{eq2}) but with a dependence on those gradients. Other investigations, such as \citep{ref42}, also worked with these weights and considered them as adaptive penalties adjustable by the neural architecture using maximum likelihood estimation, obtaining state-of-the-art results for various numerical experiments. However, all these efforts have been made for a framework where the initial and boundary conditions are imposed in a \textit{soft} manner. On the other hand, \citep{ref36} proposed an approach where the boundary conditions are imposed a priori, forcing the network to take them into account through an assembly of architectures, one of which learns the conditions and the other solves the PDEs in the $\Omega$ domain.

In terms of variations of the algorithm to improve the performance of PINNs, studies such as \citep{ref41} focused on working on the sampling method for the physical domain $[0,T]\times\Omega$ used as training data for a neural architecture. The authors proposed different types of sampling, as well as oversampling of regions with the highest accumulated residue (residual-based adaptive refinement). In other studies such as \citep{ref2}, a direct modification of the neural architecture, directly affecting the activation functions in general, was proposed. These functions (\textit{ReLU}, \textit{tanh}, \textit{sigmoid}, etc.) include the slope parameter, which can be considered as another training variable and modified in the process, leading to an improvement in convergence as well as performance on the problems encountered.

As for the solution of problems with discontinuities, some references have proposed to use the method of characteristics for problems such as the classical Euler equations. This method modifies the conservation equation (\ref{eq3}) to work as the eigenvalues and eigenvectors of the Jacobian matrix, thus exploiting the full potential of these equations (see \citep{ref19}, Section 4.1.6). There are also recent works such as \citep{ref3} that have focused on adapting the loss function to the physical problem by considering a multiplicative factor that depends on the velocity gradient in the classical Euler equations. With this premise, the authors have managed to significantly improve the algorithm in detecting discontinuities.

In this work, we present Gradient-Annihilated PINNs (GA-PINNs), an alternative approach to the methods available in the literature for resolving physical systems with discontinuities throughout their evolution. Our investigation is focused on the physical system given by the RH (Euler) equations. To the best of our knowledge, our work is the 
first one that uses PINNs to solve systems with discontinuities for the case of relativistic flows. 

\section{Methodology}
\label{sec:methods}
For our methodology, we will consider a fully connected dense neural network with a total number of layers $K$, including the input and output layers as well as the internal (hidden) ones. Each layer $k$ will consist of $N_{k}$ neurons in general. Furthermore, each layer will take a non-linear activation function, $\sigma_{k}$, which will transform the output of said layer before being transported to the subsequent layer. Mathematically, one can write the output of the $k^{\text{th}}$ layer as
\begin{equation}
\mathcal{U}_{\Theta,k}:=\sigma_{k}\left(W_{k}\mathcal{U}_{\Theta,k-1}+b_{k}\right),
\label{eq10}
\end{equation}
where $W_{k}$ and $b_{k}$ correspond to the network weights and biases of the $k^{\text{th}}$ layer.

\begin{figure}[h]
\centering
\includegraphics[scale=0.32]{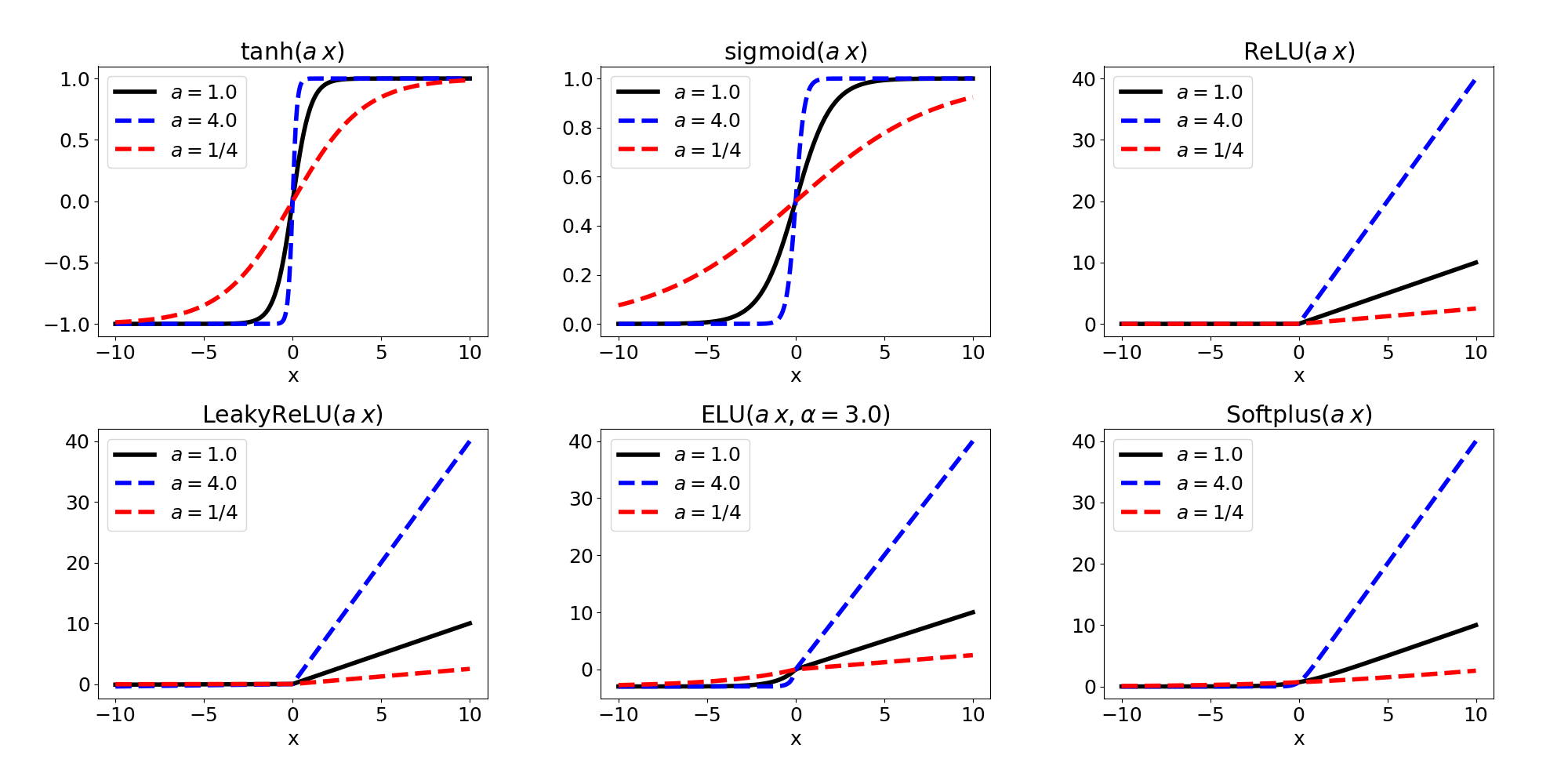} 
\caption{Different non-linear activation functions for different values of slope ($a$) accompanying the independent variable $x$.}
\label{Fig1}
\end{figure}

Hence, when additional parameters are not taken into account, the collection of modifiable quantities in a typical neural network can be represented as the combination of weights and biases, i.e. $\Theta:=\{W_{k},b_{k}\}_{1\leq k\leq K}$. With respect to activation functions ($\sigma$), it is typical for them to vary across layers and even across individual neurons. Depending on the particular problem at hand, distinct functions may be employed, such as the hyperbolic tangent or the \textit{sigmoid} function, among others. These functions, and others, have been extensively documented and utilized in literature, and are illustrated in graphical form in Figure \ref{Fig1}.
When constructing a neural network, there exists a substantial degree of flexibility in determining its dimensions, including the number of neurons per layer and the selection of the activation function employed across the various layers. These activations can contain, in turn, some hyperparameters that can be chosen a priori, i.e., before training. One could discuss the slope denoted as $a$, which works as a multiplicative coefficient for the independent variable. Nonetheless, the selection of such hyperparameters is typically achieved through an empirical approach. Thus, an extension to the PINNs algorithm has been suggested in \citep{ref2}, wherein the adaptability and trainability of the parameter $a_{k}$ for the $k^{\text{th}}$ layer are taken into consideration. In addition and as a clarification, it is also possible to consider a non-trainable constant acceleration factor $n_{k}\geq 1$ multiplying the parameter $a_{k}$ so that it can be trained faster. The aforementioned outcome could also be achieved by contemplating an elevated learning rate for the slopes relative to the other trainable parameters. Hence, by taking into account the output of the layer $k$ in equation (\ref{eq10}) and utilizing the input $(t,x)\in[0,T]\times\Omega$, where $\Omega\subset\mathbb{R}$, as the input to the network, it is feasible to predict the physical parameters as the resultant output:
\begin{equation}
\mathcal{U}(t,x)\approx\mathcal{U}_{\Theta}(t,x)=\sigma_{K}\left(\mathcal{U}_{\Theta,K}\circ\sigma_{K-1}\circ\mathcal{U}_{\Theta,K-1}\circ...\circ\sigma_{1}\circ\mathcal{U}_{\Theta,1}\right)(t,x),
\label{eq11}
\end{equation}
where $\circ$ represents the composition operator.

In equation (\ref{eq11}), each layer $k$ of the neural network is associated with an activation function $\sigma_{k}$. This function acts on each element of the corresponding output tensor, and it is possible for the activation functions to vary across layers, with adaptive and independent slopes $a_{k}$. Notably, the activation function $\sigma_{K}$ of the final output layer may be subject to additional constraints, such as ensuring non-negative or bounded values, in order to align with physical requirements. Once the activation functions have been determined, the problem of training the network can be cast as one of minimizing the loss function, which involves jointly optimizing the weights, biases, and slopes:
\begin{displaymath}
a_{k}^{*}:=\text{arg min}\left(\mathcal{L}(a_{k})\right),
\end{displaymath}
where $\mathcal{L}$ represents the loss function that can generally be written as (\ref{eq2}) for physical problems.

It should be noted that the additional trainable parameters are the slopes $\{a_{k}\}_{1\leq k\leq K}$, but considering an acceleration factor $n_{k}$, the activation function for the $k^{\text{th}}$ layer will be written as
\begin{displaymath}
\sigma_{k}:=\sigma_{k}(n_{k}a_{k}),
\end{displaymath}
where $n_{k}$ can be chosen a priori and the initialization of the slopes will be such that $n_{k}a_{k}=1$ in order to ensure convergence, since by increasing the acceleration factor $n_{k}$ the parameter $a_{k}$ becomes more sensitive.

\subsection{Modified loss function and optimization}
\label{sec:31}
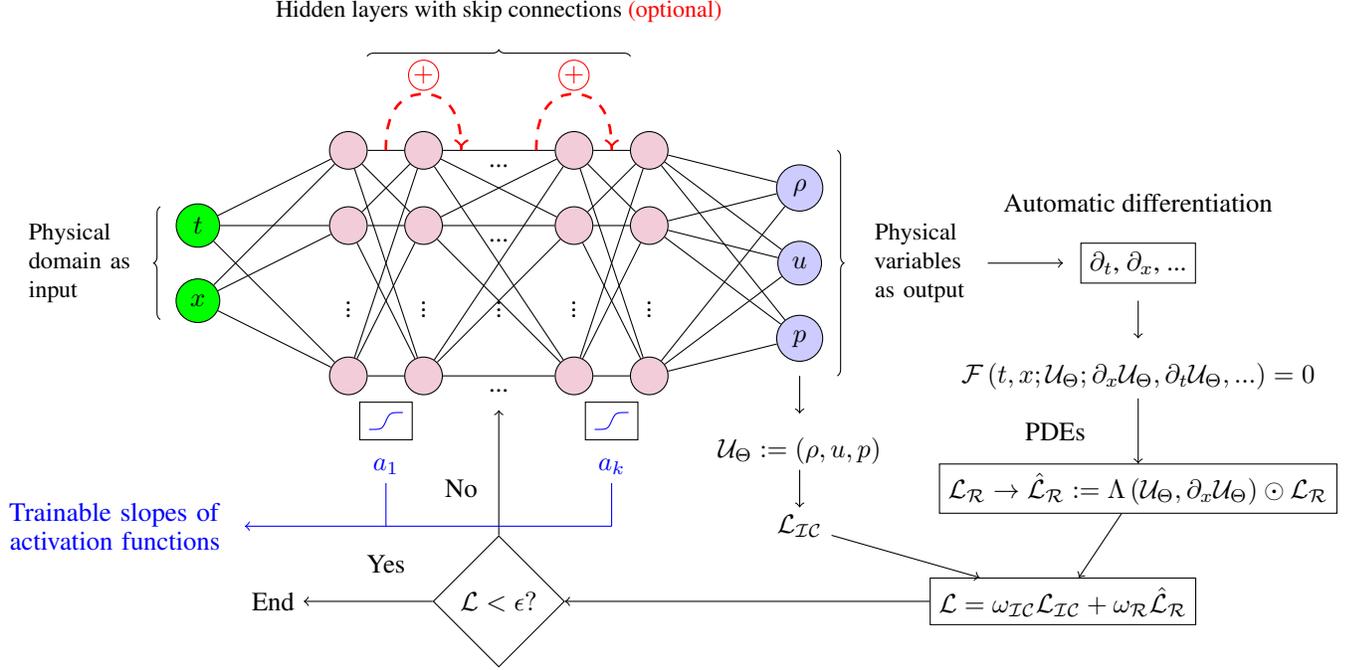
\begin{figure}[h]
\centering
\begin{tikzpicture}
\node[circle, draw=black, fill=green](t) at (0,0) {$t$};
\node[circle, draw=black, fill=green](x) at (0,-1) {$x$};

\node[circle, draw=black, fill=purple!20, minimum size=0.5cm](w1_1) at (2,1) {};
\node[circle, draw=black, fill=purple!20, minimum size=0.5cm](w2_1) at (2,0) {};
\node[minimum size=0.5cm, rotate=90, text width=0.5cm](w3_1) at (2,-1) {...};
\node[circle, draw=black, fill=purple!20, minimum size=0.5cm](w4_1) at (2,-2) {};

\node[circle, draw=black, fill=purple!20, minimum size=0.5cm](w1_2) at (3,1) {};
\node[circle, draw=black, fill=purple!20, minimum size=0.5cm](w2_2) at (3,0) {};
\node[minimum size=0.5cm, rotate=90, text width=0.5cm](w3_2) at (3,-1) {...};
\node[circle, draw=black, fill=purple!20, minimum size=0.5cm](w4_2) at (3,-2) {};

\node[minimum size=0.5cm, yshift=-0.2cm](w1_3) at (4,1) {...};
\node[minimum size=0.5cm, yshift=-0.2cm](w2_3) at (4,0) {...};
\node[minimum size=0.5cm, rotate=90, text width=0.5cm](w3_3) at (4,-1) {...};
\node[minimum size=0.5cm, yshift=-0.2cm](w4_3) at (4,-2) {...};

\node[circle, draw=black, fill=purple!20, minimum size=0.5cm](w1_4) at (5,1) {};
\node[circle, draw=black, fill=purple!20, minimum size=0.5cm](w2_4) at (5,0) {};
\node[minimum size=0.5cm, rotate=90, text width=0.5cm](w3_4) at (5,-1) {...};
\node[circle, draw=black, fill=purple!20, minimum size=0.5cm](w4_4) at (5,-2) {};

\node[circle, draw=black, fill=purple!20, minimum size=0.5cm](w1_5) at (6,1) {};
\node[circle, draw=black, fill=purple!20, minimum size=0.5cm](w2_5) at (6,0) {};
\node[minimum size=0.5cm, rotate=90, text width=0.5cm](w3_5) at (6,-1) {...};
\node[circle, draw=black, fill=purple!20, minimum size=0.5cm](w4_5) at (6,-2) {};

\node[circle, draw=black, fill=blue!20, yshift=0.5cm](rho) at (8,0) {$\rho$};
\node[circle, draw=black, fill=blue!20, yshift=-0.5cm](u) at (8,0) {$u$};
\node[circle, draw=black, fill=blue!20, yshift=-1.5cm](p) at (8,0) {$p$};

\node[](U) at (8,-3) {$\mathcal{U}_{\Theta}:=(\rho,u,p)$};
\draw[->] (8,-2) -- (8,-2.5);

\draw (t) -- (w1_1);
\draw (t) -- (w2_1);
\draw (t) -- (w4_1);

\draw (x) -- (w1_1);
\draw (x) -- (w2_1);
\draw (x) -- (w4_1);

\draw (w1_1) -- (w1_2);
\draw (w1_1) -- (w2_2);
\draw (w1_1) -- (w4_2);

\draw (w2_1) -- (w1_2);
\draw (w2_1) -- (w2_2);
\draw (w2_1) -- (w4_2);

\draw (w4_1) -- (w1_2);
\draw (w4_1) -- (w2_2);
\draw (w4_1) -- (w4_2);

\draw (w1_2) -- (w1_4);
\draw (w1_2) -- (w2_4);
\draw (w1_2) -- (w4_4);

\draw (w2_2) -- (w1_4);
\draw (w2_2) -- (w2_4);
\draw (w2_2) -- (w4_4);

\draw (w4_2) -- (w1_4);
\draw (w4_2) -- (w2_4);
\draw (w4_2) -- (w4_4);

\draw (w1_4) -- (w1_5);
\draw (w1_4) -- (w2_5);
\draw (w1_4) -- (w4_5);

\draw (w2_4) -- (w1_5);
\draw (w2_4) -- (w2_5);
\draw (w2_4) -- (w4_5);

\draw (w4_4) -- (w1_5);
\draw (w4_4) -- (w2_5);
\draw (w4_4) -- (w4_5);

\draw (w1_5) -- (rho);
\draw (w2_5) -- (rho);
\draw (w4_5) -- (rho);

\draw (w1_5) -- (u);
\draw (w2_5) -- (u);
\draw (w4_5) -- (u);

\draw (w1_5) -- (p);
\draw (w2_5) -- (p);
\draw (w4_5) -- (p);

\draw [decorate, decoration = {brace,mirror}] (-0.5,0.25) --  (-0.5,-1.25) node [black,midway,text width=1.5cm,xshift=-1cm]
{\footnotesize Physical domain as input};

\draw [decorate, decoration = {brace}] (2.25,2.25) --  (5.75,2.25) node [black,midway,yshift=0.6cm]
{\footnotesize Hidden layers with skip connections {\color{red}(optional)}};

\draw[red, line width=1pt, dashed, smooth, tension=2, dash pattern=on 4pt off 4pt, postaction={decorate,decoration={markings, mark=at position 0.98 with {\arrow[line width=1pt]{>}}}}]  plot coordinates {(2.5,1.0) (3.0,1.75) (3.5,1.0)};

\draw[red, line width=1pt, dashed, smooth, tension=2, dash pattern=on 4pt off 4pt, postaction={decorate,decoration={markings, mark=at position 0.98 with {\arrow[line width=1pt]{>}}}}]  plot coordinates {(4.5,1.0) (5.0,1.75) (5.5,1.0)};

\draw[red] (3.0,2.0) circle (0.20);
\node at (3.0,2.0) [text=red] {$+$};

\draw[red] (5.0,2.0) circle (0.20);
\node at (5.0,2.0) [text=red] {$+$};

\draw [decorate, decoration = {brace}] (8.5,1) --  (8.5,-2) node [black,midway,text width=1.5cm,xshift=1.25cm](text2)
{\footnotesize Physical variables as output};

\node[rectangle, draw=black](text3) at (12.5,-0.5) {$\partial_{t}$, $\partial_{x}$, ...};
\node[](text4) at (12.5,0.3) {Automatic differentiation};
\draw[->] (10.5,-0.5) -- (11.5,-0.5);

\node[](text4) at (12.5,-2) {$\mathcal{F}\left(t,x;\mathcal{U}_{\Theta};\partial_{x}\mathcal{U}_{\Theta},\partial_{t}\mathcal{U}_{\Theta},...\right)=0$};
\draw[->] (12.5,-1) -- (12.5,-1.5);

\node[](text5) at (8,-4) {$\mathcal{L}_{\mathcal{IC}}$};
\draw[->] (8,-3.25) -- (8,-3.75);

\node[rectangle, draw=black](text6) at (12.5,-3.5) {$\mathcal{L}_{\mathcal{R}}\rightarrow\hat{\mathcal{L}}_{\mathcal{R}}:=\Lambda\left(\mathcal{U}_{\Theta},\partial_{x}\mathcal{U}_{\Theta}\right)\odot\mathcal{L}_{\mathcal{R}}$};
\draw[->] (text4) -- (text6);

\node[text width=1cm](text7) at (11.5,-2.75) {PDEs};

\node[rectangle, draw=black](text8) at (11.5,-5) {$\mathcal{L}=\omega_{\mathcal{IC}}\mathcal{L}_{\mathcal{IC}}+\omega_{\mathcal{R}}\hat{\mathcal{L}}_{\mathcal{R}}$};

\draw[->] (text5) -- (text8);
\draw[->] (text6) -- (text8);

\node[diamond, draw=black](text9) at (4,-5) {$\mathcal{L}<\epsilon?$};
\draw[->] (text8) -- (text9);

\node[](text10) at (1,-5) {End};
\draw[->] (text9) -- (text10);
\draw[->] (text9) -- (w4_3);

\node[](text11) at (2.5,-4.5) {Yes};
\node[](text12) at (3.5,-3.5) {No};

\node [draw, minimum width=0.5cm, minimum height=0.5cm] (func1) at (2.5,-2.6) {
\begin{tikzpicture}
\draw[scale=0.075,domain=-3:3,samples=100,smooth,variable=\x,blue] plot ({\x},{3/(1+exp(-3*\x))});
\end{tikzpicture}
};

\node [draw, minimum width=0.5cm, minimum height=0.5cm] (func2) at (5.5,-2.6) {
\begin{tikzpicture}
\draw[scale=0.075,domain=-3:3,samples=100,smooth,variable=\x,blue] plot ({\x},{3/(1+exp(-3*\x))});
\end{tikzpicture}
};

\node[blue](text13) at (2.5,-3.2) {$a_{1}$};
\node[blue](text14) at (5.5,-3.2) {$a_{k}$};

\node[blue,text width=3cm](text16) at (-1.0,-4.0) {Trainable slopes of activation functions};
\draw[-,blue] (text13) -- (2.5,-4.0);
\draw[-,blue] (text14) -- (5.5,-4.0);
\draw[->,blue] (5.5,-4.0) -- (text16);

\end{tikzpicture}
\caption{General procedure of our methodology. The input to our neural network is a set of physical variables that include both time and space components, represented as the pair $(t,x)$. The neural architecture can vary in terms of the number of layers, internal activation functions, and the number of neurons per layer. Some general aspects, such as the adaptivity of the activation slopes $a_{k}$ will be dynamical, while there may be subtle variations depending on the problem at hand. The output of the network corresponds to the primitive variables $(\rho,u,p)$. These variables are subject to automatic differentiation to determine the residuals of the physical constraints and the initial conditions, which are then combined using weights $(\omega_{\mathcal{IC}},\omega_{\mathcal{R}})$. Finally, the final loss function $\mathcal{L}$ is evaluated to determine whether the procedure should terminate after updating the network parameters, $\Theta$.}
\label{NN_diagram}
\end{figure}
Despite the fact that the incorporation of adaptive and trainable activation functions can yield cutting-edge outcomes in situations featuring simple discontinuities, the majority of our methodology, the GA-PINNs, will rely on a basic alteration to the employed loss function. As indicated in (\ref{eq2}), considering the \textit{soft enforcement} of the initial and/or boundary conditions, there is a loss term that represents the fit of the physical constraints on the domain $(0,T]\times\Omega$ , i.e. $\mathcal{L}_{\mathcal{R}}$. Following the restriction in (\ref{eq1}), this part of the loss can be expressed as a MSE,
\begin{equation}
\mathcal{L}_{\mathcal{R}}:=\frac{1}{N_{\mathcal{R}}}\sum_{(0,T]\times\Omega}|\mathcal{F}|^{2},
\label{eq12}
\end{equation}
where $N_{\mathcal{R}}$ represents the number of points in the sampled internal physical domain, also called \textit{collocation points}. For its part, $\mathcal{F}$ written in (\ref{eq1}) represents our system of PDEs to be solved and in general it will depend on the physical domain $t$ and $x$ as well as on the physical fields ($\mathcal{U}$) and their derivatives with respect to space and time. In Riemann problems and in the presence of shock waves in the physical equations, any neural architecture may have difficulty resolving such discontinuities due to its tendency to solve problems smoothly. In this way, and with the aim of improving the resolution of discontinuities with PINNs, we propose the following modification of the loss of placement function (\ref{eq12}) as follows
\begin{equation}
\mathcal{L}_{\mathcal{R}}\rightarrow\hat{\mathcal{L}}_{\mathcal{R}}:=\Lambda(\mathcal{U}_{\Theta},\partial_{x}\mathcal{U}_{\Theta})\odot\mathcal{L}_{\mathcal{R}},
\label{eq13}
\end{equation}
where $\odot$ represents the element-wise product. The variable $\Lambda$ presented in equation (\ref{eq13}) is commonly considered as a function dependent on physical variables, denoted by $\mathcal{U}_{\Theta}$, and their first spatial derivatives, denoted by $\partial_{x}\mathcal{U}_{\Theta}$. In one-dimensional problems, the spatial dimension is denoted by $x$. The primary aim of $\Lambda$ is to serve as a vector of weights, providing a mixture of residuals that is distinct from what would be obtained by calculating a simple arithmetic mean. The functional form of $\Lambda$ can be arbitrary, but we shall define it in accordance with the following assumption:
\begin{center}
\begin{hypothesis}
The neural architecture increases its capacity to resolve discontinuities in PDEs when it is practically deactivated in said areas.
\end{hypothesis}
\end{center}
Which, in other words, corresponds to the fact that the PINN improves its performance (the error of its predicted solution with respect to a base resolution is lower) when the network does not pay as much attention to the discontinuous areas since it does not try to solve them smoothly. From a mathematical point of view, we can identify discontinuous areas as those that present a large gradient. Therefore, it is possible to define functions that take this into account, as
\begin{equation}
\Lambda_{1}\left(\mathcal{U}_{\Theta},\partial_{x}\mathcal{U}_{\Theta}\right):=\sum_{k=1}^{D}\frac{1}{1+\alpha_{k}\mathopen|\partial_{x}\mathcal{U}_{\Theta,k}\mathclose|^{\beta_{k}}},
\label{eq14}
\end{equation}
where $D$ corresponds to the size of the $\mathcal{U}$ vector (number of variables).

In (\ref{eq14}), $\mathcal{U}_{k}$ represents each of the predicted physical variables, i.e. $\mathcal{U}=(\rho,u,p)$ in the case of the equations in (\ref{eq9}). For their part, one can define $\bm{\alpha}:=\{\alpha_{k}\}_{k=1}^{D}$ and $\bm{\beta}:=\{\beta_{k}\}_{k=1}^{D}$ as two additional set of hyperparameters that will measure to what extent the respective spatial gradients are considered in the loss function. In particular, for the resolution of the relativistic Euler equations, (\ref{eq14}) can be particularized as:
\begin{equation}
\Lambda_{1}\left(\mathcal{U}_{\Theta},\partial_{x}\mathcal{U}_{\Theta}\right):=\Lambda_{1}(\bm{\alpha},\bm{\beta})=\frac{1}{1+\alpha_{\rho}\mathopen|\partial_{x}\rho\mathclose|^{\beta_{\rho}}}+\frac{1}{1+\alpha_{u}\mathopen|\partial_{x}u\mathclose|^{\beta_{u}}}+\frac{1}{1+\alpha_{p}\mathopen|\partial_{x}p\mathclose|^{\beta_{p}}}.
\label{eq15}
\end{equation}
From the expression in (\ref{eq15}) we can see that in areas where the gradient of the variables shoots up, i.e. in areas with discontinuities, we will have $|\partial_{x}\rho|,|\partial_{x}u|,|\partial_{x}p| \gg 1$, so that
\begin{equation}
\lim_{|\partial_{x}\rho|,|\partial_{x}u|,|\partial_{x}p|\to\infty}\Lambda_{1}(\bm{\alpha},\bm{\beta})=0.
\label{eq16}
\end{equation}
In this limiting case the part of the loss function that adjusts the physical constraints, $\mathcal{L}_{\mathcal{R}}$, would tend to be deactivated at those points in space with discontinuities. In contrast, in the opposite limit, when $|\partial_{x}\rho|,|\partial_{x}u|,|\partial_{x}p| \ll 1$, one has
\begin{equation}
\lim_{|\partial_{x}\rho|,|\partial_{x}u|,|\partial_{x}p|\to 0}\Lambda_{1}(\bm{\alpha},\bm{\beta})=3,
\label{eq17}
\end{equation}
thus recovering the PINNs base algorithm except for a constant fixed factor that can be masked within the weight $\omega_{\mathcal{R}}$. Special care should be taken in choosing the values of $(\bm{\alpha},\bm{\beta})$, but according to our hypothesis there should be at least one set of both parameters for which the error made by our algorithm with respect to a solution considered as true (e.g. analytical solution) should be smaller than or at least equal to the error obtained with an ordinary PINNs-based model without any further modification. In this way, with this simple modification it should be possible to optimize the algorithm, taking its resolving capacity to the maximum and outperforming the PINN base algorithm. Apart from the function $\Lambda$ presented in (\ref{eq15}) other expressions can also be considered as long as they present behaviours similar to those written in the limits (\ref{eq16}) and (\ref{eq17}), such as
\begin{equation}
\Lambda_{2}\left(\mathcal{U}_{\Theta},\partial_{x}\mathcal{U}_{\Theta}\right):=\Lambda_{2}(\bm{\alpha},\bm{\beta})=\frac{1}{1+\alpha_{\rho}\mathopen|\partial_{x}\rho\mathclose|^{\beta_{\rho}}+\alpha_{u}\mathopen|\partial_{x}u\mathclose|^{\beta_{u}}+\alpha_{p}\mathopen|\partial_{x}p\mathclose|^{\beta_{p}}}.
\label{eq18}
\end{equation}
The same limits for $\Lambda_{1}$ can be derived directly for $\Lambda_{2}$, obtaining
\begin{equation}
\lim_{|\partial_{x}\rho|,|\partial_{x}u|,|\partial_{x}p|\to\infty}\Lambda_{2}(\bm{\alpha},\bm{\beta})=0,\qquad\lim_{|\partial_{x}\rho|,|\partial_{x}u|,|\partial_{x}p|\to 0}\Lambda_{2}(\bm{\alpha},\bm{\beta})=1.
\label{eq19}
\end{equation}
The functions presented in equations (\ref{eq14}) and (\ref{eq18}) exhibit particular interest, as they produce weights that smoothly tend to zero in regions with discontinuities, where large gradients occur. Such behaviour is desirable, as the network should be deactivated in these areas, but not completely annihilated. Instead, the attention of the network to discontinuous regions should be negligible compared to that of smooth zones. In this way, the network can still identify the location of the discontinuities and accurately arrange the points.

\begin{figure}[h]
\centering
\includegraphics[scale=0.33]{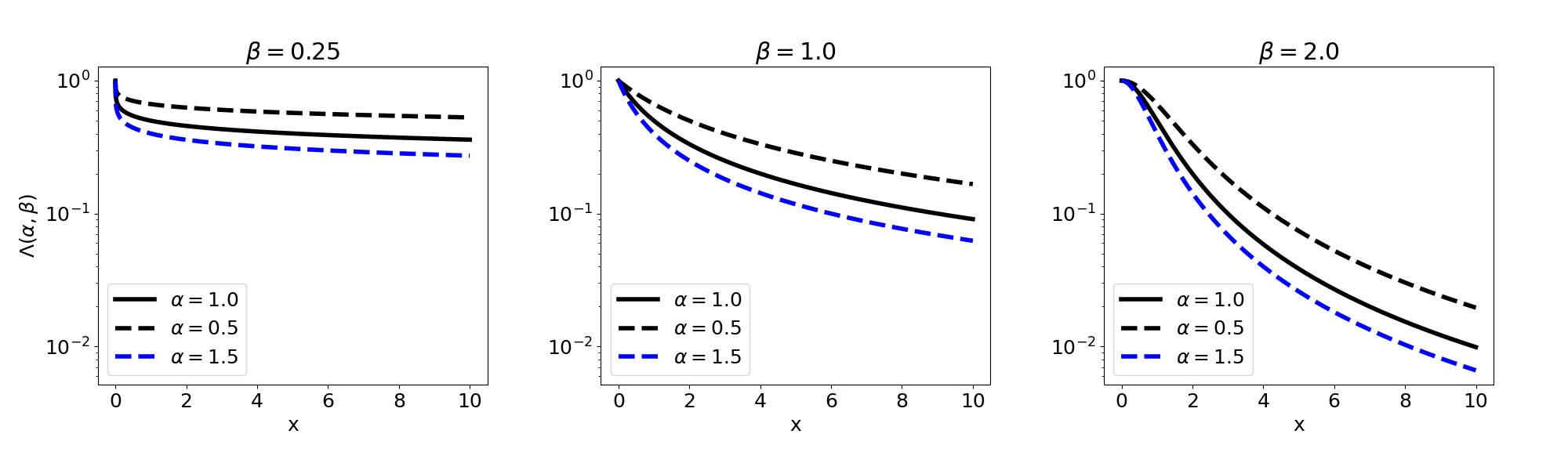} 
\caption{Example function $\frac{1}{1+\alpha\:x^{\beta}}$ against the independent variable $x$ for different values of $\alpha$ and $\beta$, all in the same scale. As both increase, the output tends to decrease in a general way.}
\label{Fig2}
\end{figure}

As a visual example, Figure \ref{Fig2} shows the function $\frac{1}{1+\alpha\:x^{\beta}}$ for different values of the hyperparameters mentioned. Observations suggest that as the parameters $\alpha$ and $\beta$ increase, the function exhibits a decreasing trend. However, it is noteworthy that a substantially greater decrease is typically observed for the latter, with differences often spanning several orders of magnitude. The underlying cause for this phenomenon is attributed to the nature of the $\beta$ parameter, which serves as an exponent to the gradient values. This non-linear transformation accentuates the disparities between the maximum and minimum values. Hence, upon scrutinizing the shape of the function $\Lambda$ as expressed in equation (\ref{eq15}), it is evident that the neural network stands to gain from the utilization of these weights. This is because the discontinuities that manifest uniformly across all variables possess $\Lambda$ values that converge towards zero. For those that are not common for all, the network might have more difficulty describing them, but as we will see, the set of hyperparameters ($\alpha,\beta$) plays an important role in these cases, together with the adaptive activation functions in both the inner and output layers. All the components converge to a competitive resolution, as it will be shown in the Results section (\ref{sec:results}).

While the set ($\bm{\alpha},\bm{\beta}$) indicates how gradients are accounted for in the loss function, the parameters ($\omega_{\mathcal{IC}},\omega_{\mathcal{R}}$) represent a mixture vector that indicates in what proportion the respective parts of the total loss function (\ref{eq2}) are to be weighed. Depending on the relative value of these hyperparameters, the network will focus on solving one of the individual losses first, thus adjusting one domain before others. In general, when working with PINNs through \textit{soft enforcement}, networks benefit from using large weights for the initial condition loss versus smaller weights for the part of the loss that tries to fit the physical constraints, i.e. $\omega_{\mathcal{IC}} \gg \omega_{\mathcal{R}}$ \citep{ref30}. It is based on the a priori knowledge that the initial conditions determine the entire evolution of the system in time $t > 0$. So by first adjusting these conditions, the network succeeds in converging faster and better to its correct temporal evolution.
\begin{algorithm}[h]
\footnotesize
\KwIn{Physical domain: $t\in[0,T]$ and $x\in\Omega\subset\mathbb{R}$ in the one-dimensional case.}
\KwOut{Set of primitive variables predicted by the NN, i.e. $\mathcal{U}_{\Theta}:=(\rho,u,p)$, in the domain.}
\textbf{Specify the training set:} Generate the domain ($t,x$) using some sampling method (e.g. \textit{Uniformly random sampling} or \textit{Sobol sequence} \citep{ref48}).\\
\textbf{Construct the neural network} as a dense fully-connected NN with $K$ layers in total and a specific number of neurons. Initialize its set of parameters, $\Theta:=\{W_{k},b_{k}\}_{1\leq k\leq K}$, in a random way.\\
\textbf{Define the activation functions}, $\sigma_{k}:=\sigma_{k}(n_{k}a_{k})$, of the layers and initialize the respective slopes, i.e. $a_{k}$, such that $n_{k}a_{k}=1$, where $n_{k}$ will be a constant and non-trainable acceleration factor. Include the set $\{a_{k}\}_{1\leq k\leq K}$ in the set of trainable parameters $\Theta$.\\
\textbf{Compute the MSE for the predicted set of variables at $(t,x)\in\{0\}\times\Omega$}:
\begin{displaymath}
\mathcal{L}_{\mathcal{IC}}=\frac{1}{N_{\mathcal{IC}}}\sum_{\{0\}\times\Omega}\mathopen|\mathcal{U}_{\Theta}(0,x)-\mathcal{U}(0,x)\mathclose|^{2}.
\end{displaymath}\\
\textbf{Calculate the spatial and temporal derivatives from the physical output variables}, $\mathcal{U}_{\Theta}$, and determine the residuals of the differential equation, $\mathopen|\mathcal{F}\mathclose|^{2}$ , for each point $(t,x)\in(0,T]\times\Omega$:
\begin{displaymath}
\mathcal{L}_{\mathcal{R}}:=\frac{1}{N_{\mathcal{R}}}\sum_{(0,T]\times\Omega}\mathopen|\mathcal{F}\mathclose|^{2}.
\end{displaymath}\\
\textbf{Calculate the function $\Lambda\left(\mathcal{U}_{\Theta},\partial_{x}\mathcal{U}_{\Theta}\right)$} and modify $\mathcal{L}_{\mathcal{R}}$ performing the product element by element:
\begin{displaymath}
\mathcal{L}_{\mathcal{R}}\rightarrow\hat{\mathcal{L}}_{\mathcal{R}}:=\Lambda\odot\mathcal{L}_{\mathcal{R}}.
\end{displaymath}\\
\textbf{Specify the final loss metric} using the mix vector $(\omega_{\mathcal{IC}},\omega_{\mathcal{R}})$:
\begin{displaymath}
\mathcal{L}=\omega_{\mathcal{IC}}\mathcal{L}_{\mathcal{IC}}+\omega_{\mathcal{IC}}\hat{\mathcal{L}}_{\mathcal{R}}.
\end{displaymath}\\
\textbf{Update the set of network variables} by backpropagation, $\{W_{k},b_{k},a_{k}\}_{1\leq k\leq K}$, that minimizes the total loss using a certain defined optimizer:
\begin{displaymath}
\Theta^{*}:=\text{arg min}\left(\mathcal{L}(\Theta)\right).
\end{displaymath}
\caption{Gradient-Annihilated PINN algorithm for the relativistic Euler problem (\ref{eq8})}
\label{algorithm1}
\end{algorithm}
By considering all the relevant factors, it is feasible to construct the algorithm of our GA-PINN methodology, which assists us both visually, as shown in Figure (\ref{NN_diagram}), and in writing, through the step-by-step algorithm (\ref{algorithm1}). In the initial stages, it is critical to create the training data. In contrast to standard machine learning models, where the complete dataset can be divided into training and testing data, both of which contain certain features that the model perceives and a label (target) that it endeavours to forecast, in PINNs, the training data are the points in the physical domain $(t,x)\in[0,T]\times\Omega$ itself. Prior to any further development, this step is vital. In general, domain sampling does not consider the creation of a test data set unless the goal is to make a prediction at a specific later point in time, for example (e.g. using LSTMs). Within our framework, it is viable to utilize the $L^{2}$ norm as a means of comparing an existing solution, referred to as the ``ground truth'' (such as an analytical solution), with the ultimate prediction produced by the neural architecture, denoted as $\mathcal{U}_{\Theta}(t,x)$.

Consequently, it can be hypothesized beforehand that the point sampling technique can have a significant influence on the effectiveness of our architecture. Nevertheless, as the size of the physical domain increases to a considerable extent, this factor may no longer be decisive. This is a crucial consideration because we aim to minimize any dependence of our findings on factors like the seed utilized in random sampling. Therefore, scenarios where the number of points tends towards infinity are preferable. In Section~\ref{sec:construction} we give a slightly longer description of some of the different sampling methods that can be used, but throughout our work we have focused on the Sobol sequence method \citep{ref48}, which fills the domain in a very uniform way. Once the domain $(t,x)$ is created, it takes on the role of input to the neural model, passes through all internal layers and comes out of said space as a prediction for the physical variables, i.e. $\mathcal{U}_{\Theta}(t,x)$. At this point, automatic differentiation methods are used to compute the temporal and spatial derivatives of the output variables. Afterwards, the respective parts of the loss function are calculated as an MSE by carefully applying the weights computed using $\Lambda$ (\ref{eq14}) to be propagated back through all the layers of the neural architecture and update the parameter set $\{W_{k},b_{k},a_{k}\}_{1\leq k\leq K}$ per gradient descent. This process is repeated for each epoch until the model converges.

\subsection{Construction of physical domain: sampling methods}
\label{sec:construction}
\begin{figure}[h]
\centering
\includegraphics[scale=0.55]{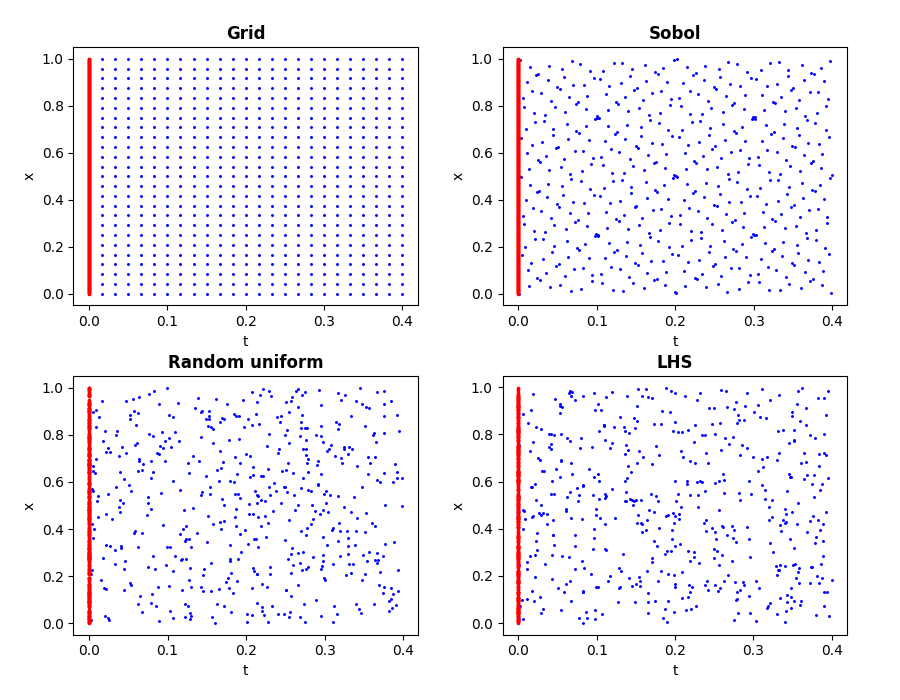} 
\caption{Examples of 500 points generated in the space $[0,0.4]\times[0,1]$ for different sampling methods. The initial domain points are shown in red, i.e. points for $t=0$, and the collocation or internal points of the domain on which the residual of the PDEs system will be calculated are shown in blue.}
\label{Fig3}
\end{figure}
PINN algorithms need to have the physical space as input for each of the problems that arise. As we have commented in Section \ref{sec:PN}, in general, this will be made up of several subdomains that will mainly attend to each of the terms of the final loss function (\ref{eq2}). In its entirety, it can be expressed mathematically as $(t,x)\in[0,T]\times\Omega$,
where $\Omega\subset\mathbb{R}$ in one-dimensional problems as the ones analysed in this paper, and where the temporal dimension runs from the initial time to a some arbitrary final instant $T$ that will depend on the problem presented. In our case, this physical space is set from start to finish for each one of the results to be presented, with the random seed set from the beginning so that all training sessions are fully repeatable and reproducible. It should be noted that despite the fact that there is some literature on the use of adaptive methods to sample the domain such as the residual-based adaptive refinement (RAR) method so that physical space is altered during training, we will not consider an adaptive method, but will focus exclusively on the effect of our presented methodology and the extent to which the loss function has resolving power.

Regarding the sampling method used, this is a somewhat arbitrary decision that is usually taken from the initial point of the work and which, for consistency, remains fixed for all the results that can be obtained. Sampling methods such as the fixed grid of uniformly distributed points or uniform sampling of random points have been widely used in studies dealing with PINNs and their variations, regardless of the nature of the physical problem. As a visual aid, we show examples of different methods in Figure \ref{Fig3}, with the uniform grid and random sampling forming the left-hand column. In the right column we present the Sobol sampling and the Latin hypercube sampling (LHS) \citep{ref49,ref50} which is a statistical method used to generate quasi-random samples using Monte Carlo by intervals using as a base the same probability and Normal distribution for each one. The Sobol sequence method has been selected for implementation in our algorithm, owing to its documented high yields in fixed sampling within the literature \citep{ref41}. This sampling method generates points in powers of two in a quasi-random manner, providing domains with relatively low discrepancy and, therefore, achieving a physical space with less overlapping of points and fewer areas without representation, but without neglecting the intrinsic randomness of the method.

\section{Numerical experiments and results}
\label{sec:results}
In this section we present the results for the solution of the special RH  equations in one spatial dimension, given by  (\ref{eq8}), considering different initial conditions. As mentioned above, solving those equations is of great interest in some areas of physics, such as high-energy physics (e.g.~to model heavy-ion collisions) and astrophysics (e.g.~to study gamma-ray bursts or the propagation of jets in active galactic nuclei). These equations are solved given some initial conditions, which are known a priori as data. Helping us from~\citep{ref1,ref5} the numerical experiments we consider are Riemann problems (i.e.~initial value problems with discontinuous data) for which there is no need to specify spatial boundary conditions. For all cases, the same physical space of size $N_{\mathcal{R}}=N_{\mathcal{IC}}=2^{15}$ points has been considered (in base 2 since we use Sobol sampling). Additionally, during all training processes, the AdamW optimizer was utilized \citep{ref51}, with an initial learning rate ranging from approximately $10^{-5}$ to $10^{-6}$, depending on the specific problem. This learning rate was intentionally set to be significantly low, as it promotes the convergence of the optimization process. For specific parameters we refer to Table \ref{Table_Global_Results_1} which reports, among other things, the weights used for the initial conditions ($\omega_{\mathcal{IC}}$) and for the PDE ($\omega_{\mathcal{R}}$) as well as the activation functions used.
\begin{table}[t]
\centering
\scriptsize

\begin{tabular}{c|ccc|ccc|c|ccc|c|ccc|}

\multicolumn{1}{c}{} & \multicolumn{1}{c}{\textbf{Hidden Layers}} & \multicolumn{1}{c}{\textbf{Neurons}} & \multicolumn{1}{c}{\bm{$\omega_{\mathcal{R}}$}} & \multicolumn{3}{c}{\bm{$\omega_{\mathcal{IC}}$}} & \multicolumn{1}{c}{\textbf{Act. hidden}} & \multicolumn{3}{c}{\textbf{Act. output}} & \multicolumn{1}{c}{\bm{$n_{\text{hidden}}$}} & \multicolumn{3}{c}{\bm{$n_{\text{output}}$}} \\ [0.25cm]
\cline{5-7}
\cline{9-11}
\cline{13-15}
\rule{0pt}{0.3cm}\textbf{Set of ICs}   &      &      &      &   $\omega_{\mathcal{IC}}^{\rho}$  &   $\omega_{\mathcal{IC}}^{u}$  &   $\omega_{\mathcal{IC}}^{p}$   &      &    $\sigma_{\rho}$  &   $\sigma_{u}$  &   $\sigma_{p}$  &       &   $n_{\rho}$  &   $n_{u}$  &   $n_{p}$\\ [0.1cm]
\cline{1-15}
\rule{0pt}{0.4cm}Problem 1  &   10  &   40  &   0.1 &   10    &   10 &   10 &   $T(x)$   &   $S(x)$  &   $S(x)$  &   $S(x)$  &   10    &   10   &   10   &   10 \\ [0.25cm]
\cline{2-15}
\rule{0pt}{0.4cm}Problem 2  &   10  &   40  &   0.1 &   10    &   10 &   10 &   $T(x)$   &   $SP(x)$  &   $T(x)$  &   $SP(x)$  &   10    &   $\varnothing$   &   10   &   $\varnothing$ \\ [0.25cm]
\cline{2-15}
\rule{0pt}{0.4cm}Problem 3  &   10  &   40  &   0.1 &   10    &   10 &   10 &   $T(x)$   &   $S(x)$  &   $T(x)$  &   $S(x)$  &   10    &   $\varnothing$   &   10   &   $\varnothing$ \\ [0.25cm]
\cline{2-15}
\rule{0pt}{0.4cm}Problem 4  &   10  &   40  &   0.1 &   100    &   10 &   10 &   $T(x)$   &   $SP(x)$  &   $T(x)$  &   $SP(x)$  &   10    &   1   &   1   &   1 \\ [0.25cm]
\cline{2-15}
\end{tabular}
\vspace{0.25cm}

\caption{Summary of the various configurations used for training our methodology for different problems. The initial condition weight is further divided into three sub-weights that correspond to different variables: $\omega_{\mathcal{IC}}^{\rho}$, $\omega_{\mathcal{IC}}^{u}$, and $\omega_{\mathcal{IC}}^{p}$. The activation of the output layer is split by variable, as is the acceleration factor, denoted by $n$, where $n_{\text{hidden}}$ corresponds to the common factor for the internal layer activations. Additionally, the notation $T(x)$, $S(x)$, and $SP(x)$ represent the hyperbolic tangent, \textit{sigmoid}, and \textit{softplus} functions, respectively. The symbol $\varnothing$ indicates that the corresponding parameter is not considered in the training process (it is fixed to 1 and not trained through the process).}
\label{Table_Global_Results_1}
\end{table}

To assess the accuracy of our predictions, we commonly employ the $L^{2}$ norm or error, which computes the squared disparities between the forecasted solution given by the constructed model and a ``ground truth'' solution. In our case, we consider the analytical solution of the Riemann problems as the ground truth, as it is readily obtainable \citep{ref5}. Furthermore, since we have more than one variable to predict and in general they can present quite different ranges of values, it is more consistent to use the relative $L^{2}$ norm and thus be able to make direct comparisons between them. The relative $L^{2}$ norm of an output variable $\mathcal{U}_{\Theta,k}$ at a certain time $t$ is then defined as
\begin{equation}
l^{2}_{\mathcal{U}_{k}}:=\sqrt{\frac{\sum_{n}^{N}\mathopen|\mathopen|\mathcal{U}_{\Theta,k}(t,x_{n})-\mathcal{U}_{k}(t,x_{n})\mathclose|\mathclose|_{2}}{\sum_{n}^{N}\mathopen|\mathopen|\mathcal{U}_{\Theta,k}(t,x_{n})\mathclose|\mathclose|_{2}}},
\label{l2_rel}
\end{equation}
where subscript $\Theta$ corresponds to the solution obtained by the model, and the summations are done along the spatial dimension. 

\begin{table}[t]
\centering
\scriptsize

\begin{tabular}{c|ccc|c|ccc|c|ccc|}

\multicolumn{1}{c}{}    &   \multicolumn{3}{c}{\textbf{GA-PINN}} & \multicolumn{1}{c}{} & \multicolumn{3}{c}{\textbf{Base-PINN}} & \multicolumn{1}{c}{} & \multicolumn{3}{c}{\textbf{HRSC models}} \\ [0.25cm]
\cline{2-4}
\cline{6-8}
\cline{10-12}
\rule{0pt}{0.3cm}\textbf{Set of ICs}   &   $l^{2}_{\rho}$  &   $l^{2}_{u}$  &   $l^{2}_{p}$   &      &    $l^{2}_{\rho}$  &   $l^{2}_{u}$  &   $l^{2}_{p}$  &       &   $l^{2}_{\rho}$  &   $l^{2}_{u}$  &   $l^{2}_{p}$\\ [0.1cm]
\cline{1-4}
\cline{6-8}
\cline{10-12}
\rule{0pt}{0.4cm}Problem 1  &   0.013    &   0.010 &    0.008   &      &   0.024  &   0.048  &   0.026  &       &   0.012   &   0.025   &   0.013 \\ [0.25cm]
\cline{2-4}
\cline{6-8}
\cline{10-12}
\rule{0pt}{0.4cm}Problem 2  &   0.137    &   0.127 &   0.029 &      &   0.351  &   0.287  &   0.053  &       &   0.073   &   0.036   &   0.013 \\ [0.25cm]
\cline{2-4}
\cline{6-8}
\cline{10-12}
\rule{0pt}{0.4cm}Problem 3  &   0.032    &   0.033 &   0.012 &      &   0.038  &   0.032  &   0.014  &       &   0.038   &   0.055   &   0.042 \\ [0.25cm]
\cline{2-4}
\cline{6-8}
\cline{10-12}
\rule{0pt}{0.4cm}Problem 4  &   0.024    &   0.043 &   0.041 &      &   0.097  &   0.197  &   0.146  &       &   0.031   &   0.070   &   0.044 \\ [0.25cm]
\cline{2-4}
\cline{6-8}
\cline{10-12}
\end{tabular}
\vspace{0.25cm}

\caption{$L^{2}$ norm relative errors of our results against the exact solution for the experiments considered. The errors are presented in sets of three boxes, where each column in each set represents a fluid variable (density, velocity, and pressure). Both the GA-PINN methodology and the base PINN model are considered (first two boxes) alongside HRSC numerical schemes \citep{ref38} (third box).}
\label{Table_Global_Results_2}
\end{table}

The following subsections present the specific results for each problem. The performance evaluation of the  models is done using the $l^{2}$ relative error metric defined in equation (\ref{l2_rel}). An overview of the results is reported in Table \ref{Table_Global_Results_2} which summarizes the values of the above metric for each of the scenarios. The errors reported correspond to the density, velocity, and pressure. Notice that each box in this table shows the results obtained with a different method, GA-PINN, Base-PINN, and HRSC schemes, which allows for a direct comparison of the methods. As it can be seen from the table, the relative errors of our GA-PINN method (in the left column) are smaller than those of the baseline PINN method in general (reported in the middle column). Moreover, our method gives competitive results when compared with those from HRSC schemes (shown in the right column). Remarkably, our model provides a significant improvement over the baseline PINN approach. The only exception is found for Problem 3, where the improvement is comparatively smaller.

\begin{figure}[t]
\centering
\includegraphics[scale=0.50]{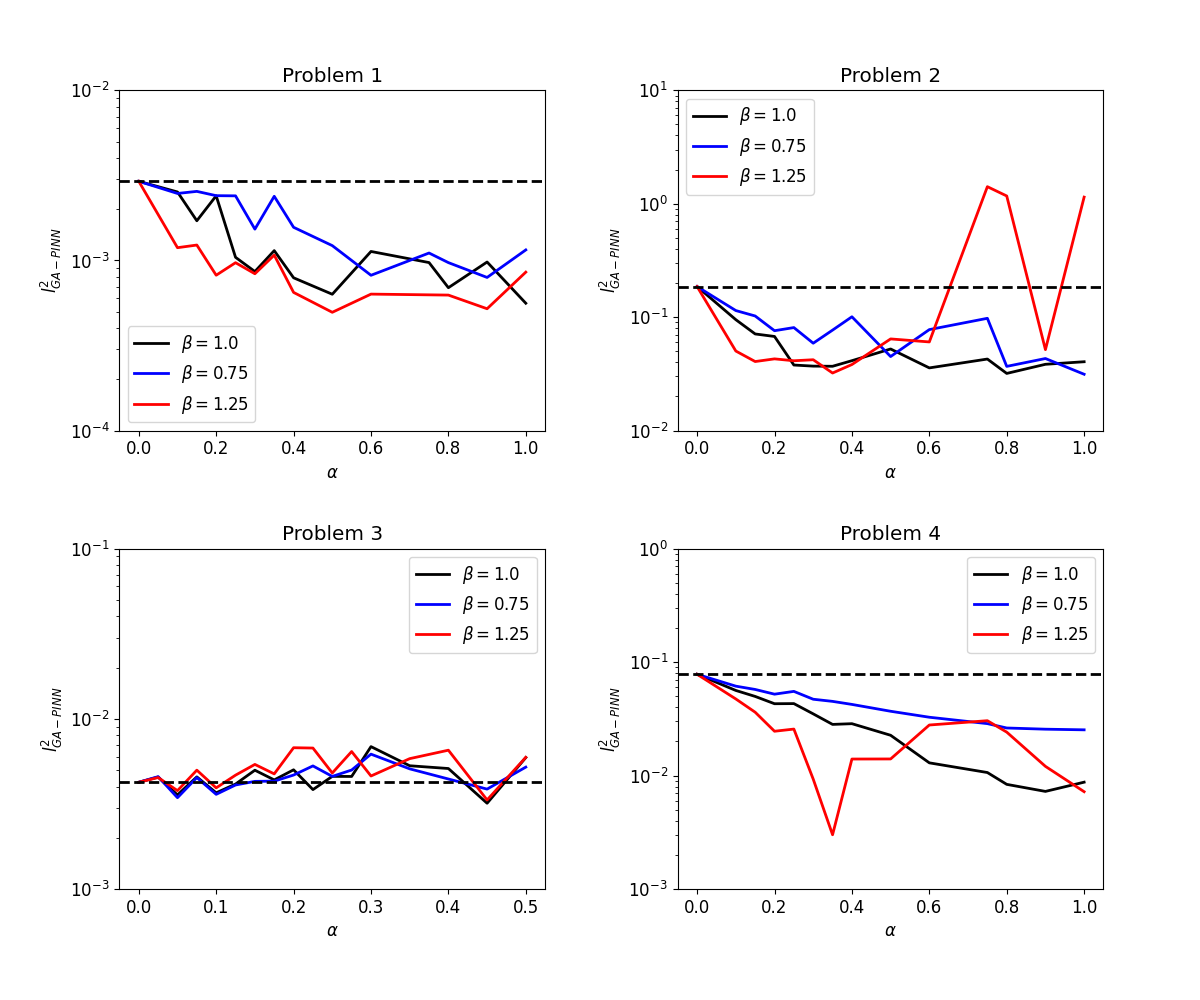} 
\caption{Relationship between the variation of the relative error $l^{2}$ for the model GA-PINN (solid lines) and the corresponding variations of the model hyper-parameters $\alpha$ and $\beta$ for a set of problems assumed to be the same for all variables for simplicity. The dashed black line in each plot corresponds to the baseline PINN model. In general, the error decreases as the weight of the gradient ($\alpha$) increases. Each point in these plots corresponds to the final result of an independent training of 40,000 epochs with a fixed learning rate of $10^{-5}$.}
\label{General_Results_L2_vs_Alpha}
\end{figure}

The results will be analysed from different perspectives. On the one hand, we will deal with borderline scenarios where the proposed neural model takes advantage of the disappearance of the gradient. On the other hand, there will also be special cases where the solutions obtained with different methods are remarkably similar.
Figure \ref{General_Results_L2_vs_Alpha} shows the variation of the $l^{2}$ error with hyperparameter $\alpha$, which is the physical gradient weight, and with exponent $\beta$ \footnote{The parameter sets $\bm{\alpha}=\{\alpha_{k}\}_{k=1}^{D}$ and $\bm{\beta}=\{\beta_{k}\}_{k=1}^{D}$ have been condensed in the figure into two unique hyperparameters for the sake of simplicity, independent of the physical variable.}. This figure allows to evaluate the hypothesis proposed in Subsection \ref{sec:31} and motivating our study, i.e.~that neural networks based on the PINN algorithm use their near-disruption in most physical domains dominated by discontinuities to improve their resolution capabilities in such regions. The reduction in the error shown in Figure \ref{General_Results_L2_vs_Alpha} is evident when the physical gradient is weighted more heavily, with the error of a baseline PINN model ($\alpha=\beta=0$) shown as a dashed line. It is apparent that the resolution capabilities of the model improve in all scenarios except in Problem 3, where it is limited to constructing noise around the discontinuous error (see below). This result is a direct consequence of our previous observations, to which we will refer to in Subsection \ref{sec:problem3}. On the other hand, increasing $\beta$ typically has a direct impact on the stability of the error as the importance of the gradient increases. This can be observed, for instance, in the plot corresponding to Problem 2 (top right panel in Figure \ref{General_Results_L2_vs_Alpha}), where a pronounced instability occurs when $\beta=1.25$ for values above $\alpha=0.5$. This behaviour is also discernible in Problem 4, although to a lesser degree. The underlying cause of this effect is because the hyperparameter $\beta$ corresponds to the exponent of the gradients, and slight variations in their values can result in substantially different outcomes. Therefore, it is essential to exercise caution when selecting this parameter, with $\beta=1.0$ constituting a consistent choice.

\subsection{Problem 1: Sod Shock Tube}
\label{sec:problem1}
The first case we  consider is a modification of a problem that has been extensively discussed in the literature on solving PDEs with PINNs: the \textit{Sod Shock Tube} problem \citep{ref29}. 
The initial data in the original problem involves a discontinuous reduction in density and pressure along the spatial dimension, while maintaining a constant zero velocity in the entire domain.  However, in our setup we incorporate a non-zero velocity to leverage equations  (\ref{eq8}) for  solving this problem. Our analysis employs an adiabatic index of $\Gamma=4/3$. Considering the physical space   $(t,x) \in [0,0.2]\times[0,1]$ and taking $x=0.5$ as the point where the discontinuities are located at $t=0$, the initial conditions for this problem are as follows:

\begin{equation}
\mathcal{U}(0,x):=\mathcal{U}_{0}(x)
=\left\{
\begin{array}{ll}
\rho_{0}(x)=1.0\quad\text{if $\,x\leq 0.5$,}\quad\rho_{0}(x)=0.125\quad\text{otherwise}\\
u_{0}(x)=0.5\quad\forall x\\
p_{0}(x)=1.0\quad\text{if $\,x\leq 0.5$,}\quad p_{0}(x)=0.1\quad\text{otherwise}\\
\end{array} 
\right.
\label{ICs_problem0}
\end{equation}

\begin{figure}[h]
\centering
\includegraphics[scale=0.55]{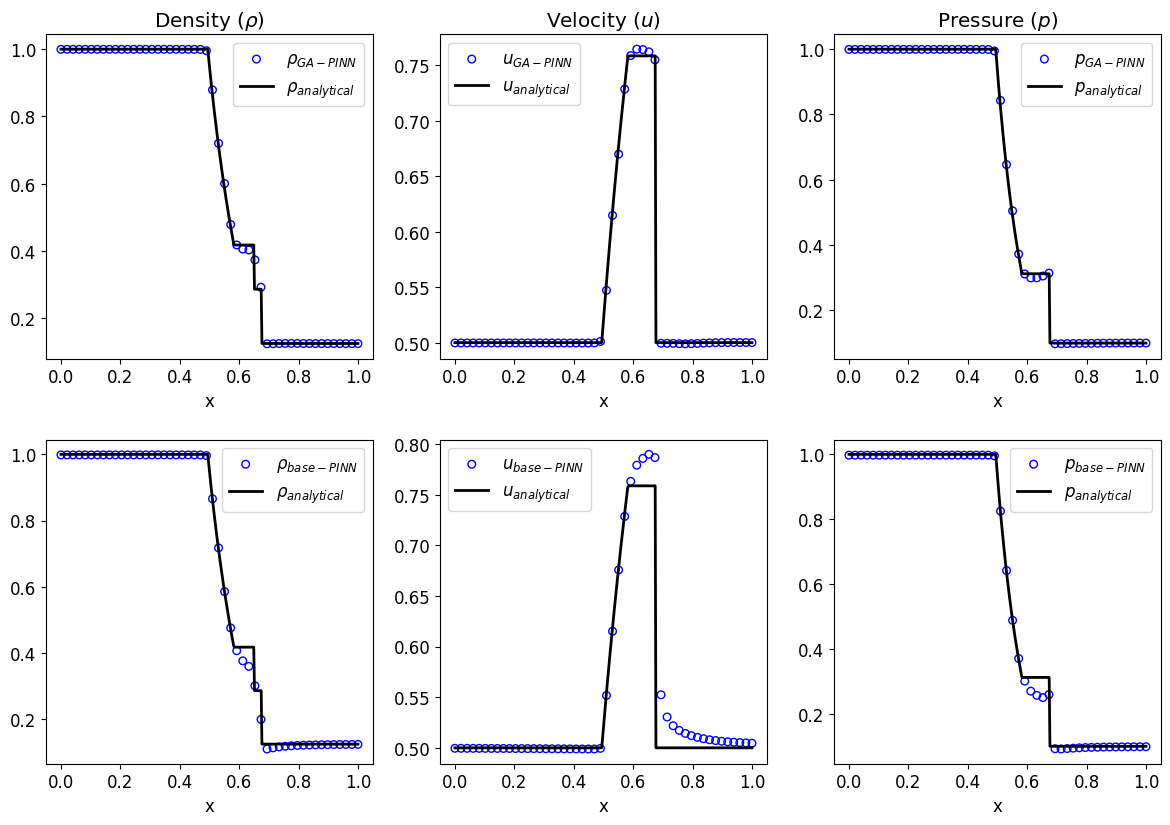} 
\caption{Solution for the physical variables for the initial conditions of Problem 1, described in Eq.~(\ref{ICs_problem0}). The results correspond to the final time $t=0.2$. The analytical solution, which serves as the reference standard, is represented by a solid black curve in all plots. The solutions obtained with our GA-PINNs method are shown in the top row while those obtained with a base PINN model are displayed in the bottom row. In the two cases the solutions are represented by blue circles.}
\label{Fig_Results_Problem_1}
\end{figure}

The evolution of this Riemann problem leads to the appearance of three elementary waves, a left-moving rarefaction fan, a right-moving shock wave and a right-moving contact discontinuity between the two. The solution at $t=0.2$ is plotted in Figure \ref{Fig_Results_Problem_1}, which shows the results for the set of primitive variables $(\rho,u,p)$. The shock wave, visible in all three variables, is located at $x=0.674$, while the contact discontinuity, only visible in the density, is at $x=0.649$. The three plots at the top row of the figure display the predictions of our method, while those at the bottom show the solution obtained with a baseline PINN model without additional modifications. For all three variables, the results for both methods are shown with blue open circles using as basis for the comparison of the analytical solution of this Riemann problem, depicted with a solid black line in each plot. Furthermore, both trainings are identical in all common parameters, including the number of training epochs. As it can be seen, the solution obtained with our method satisfactorily reproduces the correct evolution of this problem and, in particular, sharply captures both the shock wave and the contact discontinuity. This is in contrast to the base PINN model, which performs poorly. The base model is inadequate to accurately capture the contact discontinuity in density ($\rho$) and the shock wave in the velocity ($u$) due to the extended spatial gradient and the tendency to produce smooth solutions.  Furthermore, although the base model correctly localizes the position of the shock wave, it yields incorrect results for the constant state between the tail of the rarefaction wave and the shock visible both in the velocity and the pressure. As the top row of Figure \ref{Fig_Results_Problem_1} shows, our new technique mitigates all of these issues. 

\begin{figure}[h]
\centering
\includegraphics[scale=0.45]{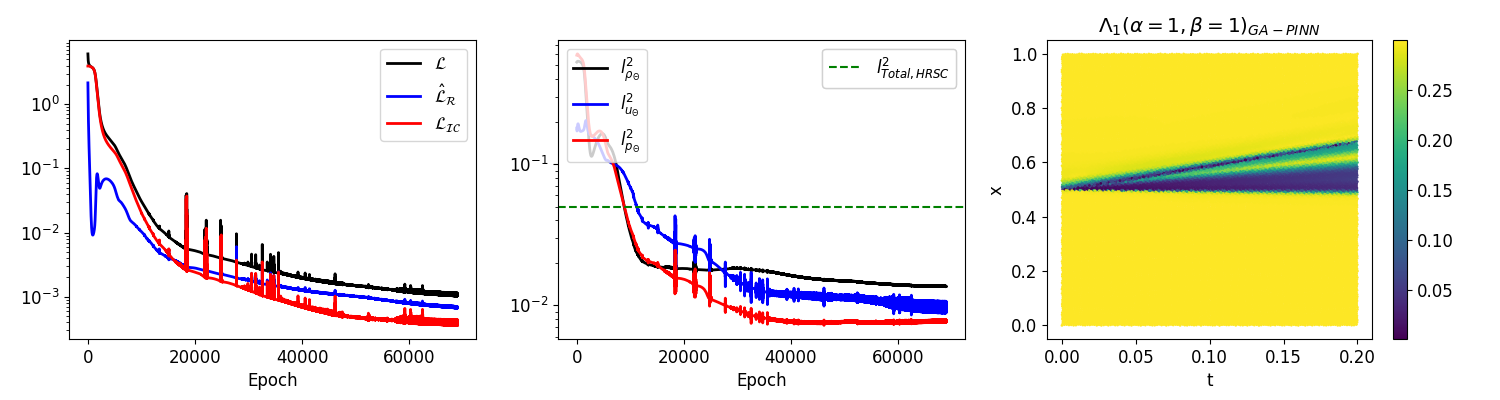} 
\caption{Evolution of the physical losses and relative errors with the epoch number for Problem 1. The physical losses $\hat{\mathcal{L}}_{\mathcal{R}}$, $\mathcal{L}_{\mathcal{ IC }}$, and the total loss $\mathcal{L}$ are shown on the left plot. The middle plot displays the relative  $l^{2}$ errors between the solution obtained with our method and the analytical solution (``ground truth''). The dashed green line represents the corresponding error for the numerical solution obtained using the second-order central HRSC scheme of~\cite{ref1} with a mesh of 400 zones ($\Delta x=0.0025$). Finally, the heat map on the right plot shows the $\Lambda$ function after training.}
\label{Fig_Results_Problem_1_extra}
\end{figure}

The function $\Lambda$ used in this problem is that given by  equation (\ref{eq15}), setting the values of $\alpha$ and $\beta$ to $\alpha_{\rho}=\alpha_{u}=\alpha_{p}=1$ and $\beta_{\rho}=\beta_{u}=\beta_{p}=1$, respectively. This selection of parameters yields results that are commensurate with those obtained via HRSC approaches. The choice of these hyperparameters is initially arbitrary, although it is possible to carry out a sweep and compute the variation in the performance of the models based on them. However, our choice is quite standard and not too aggressive, assuming a simple and direct variation of the base PINN model. The   final $\Lambda$ function we obtain is displayed on the right plot of Figure \ref{Fig_Results_Problem_1_extra}. This figure shows a heat map in physical space $(t,x)$, where colours close to yellow indicate the areas with the highest values of $\Lambda$ and colours close to blue the lowest values. Those correspond, respectively,  to very smooth areas (with small gradients) and areas with significant changes (with high gradients). One can observe a predominant blue zone whose onset is at about $x=0.5$ corresponding to the expansion zone of the fluid. Moreover, one can also see a more concentrated blue zone above $x=0.6$, which corresponds to the shock wave discontinuity in Figure \ref{Fig_Results_Problem_1}.
\begin{figure}[h]
\centering
\includegraphics[scale=0.4]{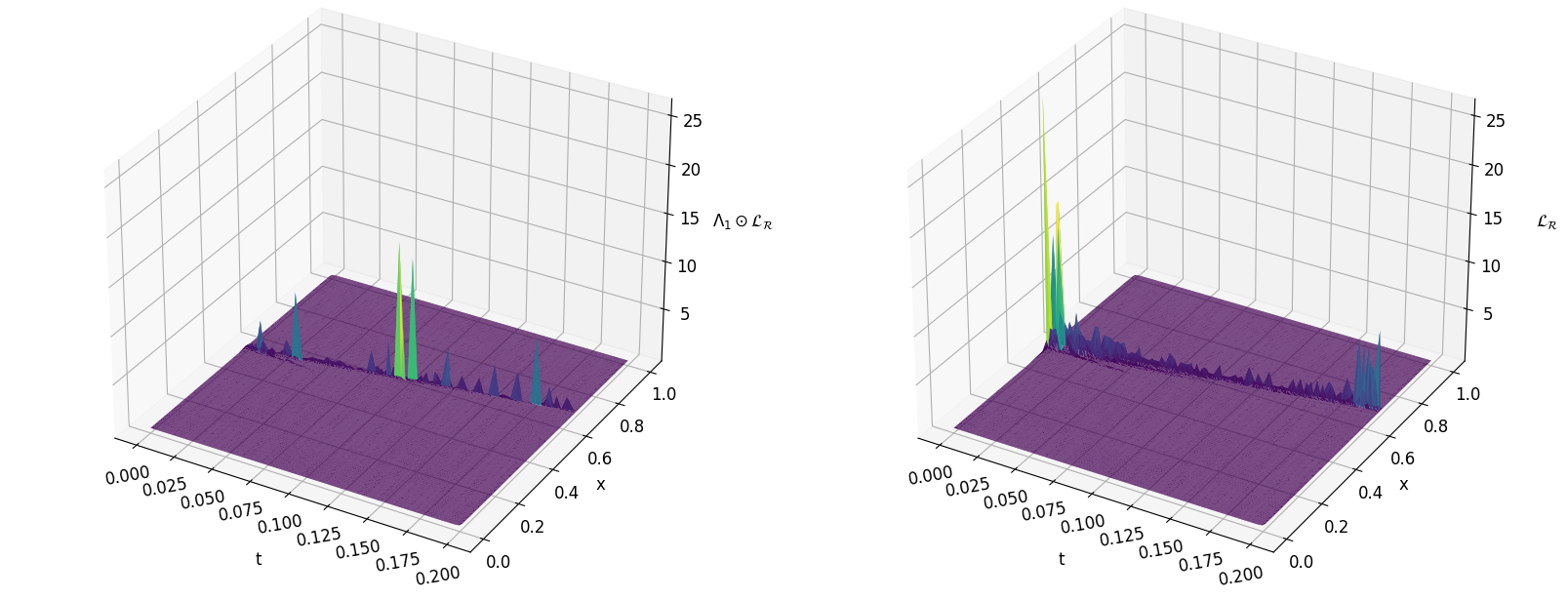} 
\caption{Three-dimensional representation of the physical losses used for the neural models. On the left is the one used for training the GA-PINN, $\Lambda_{1}\odot\mathcal{L}_{\mathcal{R}}$, while on the right is the one used for training a base PINN model. The physical errors are significantly smaller for our method, especially at times close to the initial time.}
\label{3D_plots_Problem_1}
\end{figure}

The other two plots of  Figure \ref{Fig_Results_Problem_1_extra} correspond to the total physical loss used to train the neural model and its breakdown in the loss of the initial conditions, $\mathcal{L}_{\mathcal{IC}}$, and the PDE itself, $\hat{\mathcal{L}}_{\mathcal{R}}$ (\ref{eq13}) (left plot), as well as the relative $l^{2}$ errors obtained for each of the variables (middle plot). The good behaviour in the loss during the training process is noticeable both in the left plot and in the relative error for each variable computed according to (\ref{l2_rel}) displayed in the middle plot of the figure. The total value of the latter is $l^{2}_{\text{GA-PINN}}=0.031$ which we can compare to a value of $l^{2}_{\text{HRSC}}=0.050$ obtained with the HRSC central scheme of \cite{ref1} (green dashed line), as specified in Table \ref{Table_Global_Results_2}, being slightly above only for the density. 
This comparison shows that 
the proposed methodology allows us to appropriately capture the location and speed of the different waves with a precision comparable to that obtained with HRSC methods. 

It is also possible to analyse the physical errors obtained as a function of time and space by representing them in three dimensions, i.e. $\hat{\mathcal{L}}_{\mathcal{R}}(t,x)=\Lambda\odot\mathcal{L}_{\mathcal{R}}(t,x)$, considering only the one-dimation points. Figure \ref{3D_plots_Problem_1} shows these losses for both our methodology (left) and a baseline PINN model (right), using the same scale in the two cases. The first  observation is that the physical errors for the GA-PINN are not only lower in the final time ($t=0.2$) and in the initial times, but in general for the whole range. It is worth mentioning the accumulation of errors around $t=0$ for the baseline PINN model, a region where our method improves significantly. This is a great help for the neural network, as the initial conditions are the only points known a priori, and they determine the overall behaviour of the fluid for $t > 0$. In this way, correctly learned and defined initial conditions are a guarantee for the correct propagation in the entire time domain.

\subsection{Problem 2}
\label{sec:problem2}

As a second example we study a problem in which the density of the fluid remains initially constant throughout the spatial domain while the pressure has a discontinuity of an order of magnitude. In addition, half of the fluid has a certain constant speed at the initial instant while the other half is at rest. The spacetime domain considered for our solution is $(t,x)\in[0,0.4]\times[0,1]$ and the adiabatic index of the fluid is $\Gamma=4/3$. The particular initial conditions  read\footnote{The initial conditions proposed for the different problems may be normalized to the interval $[0,1]$ before the training by explicitly dividing their values on both sides by the largest value, which in this particular case is the pressure for $x>0.5$, i.e. $p_{R}=10.0$. By doing so, it is possible to continue using as activation functions usual functions such as the \textit{sigmoid} function or the hyperbolic tangent without the need to rescale them.}:

\begin{equation}
\mathcal{U}_{0}(x)
=\left\{
\begin{array}{ll}
\rho_{0}(x)=1.0\quad\forall x\\
u_{0}(x)=0.9\quad\text{if $x\leq 0.5$,}\quad u_{0}(x)=0.0\quad\text{otherwise}\\
p_{0}(x)=1.0\quad\text{if $x\leq 0.5$,}\quad p_{0}(x)=10.0\quad\text{otherwise}\\
\end{array} 
\right.
\label{ICs_problem1}
\end{equation}

\begin{figure}[t]
\centering
\includegraphics[scale=0.55]{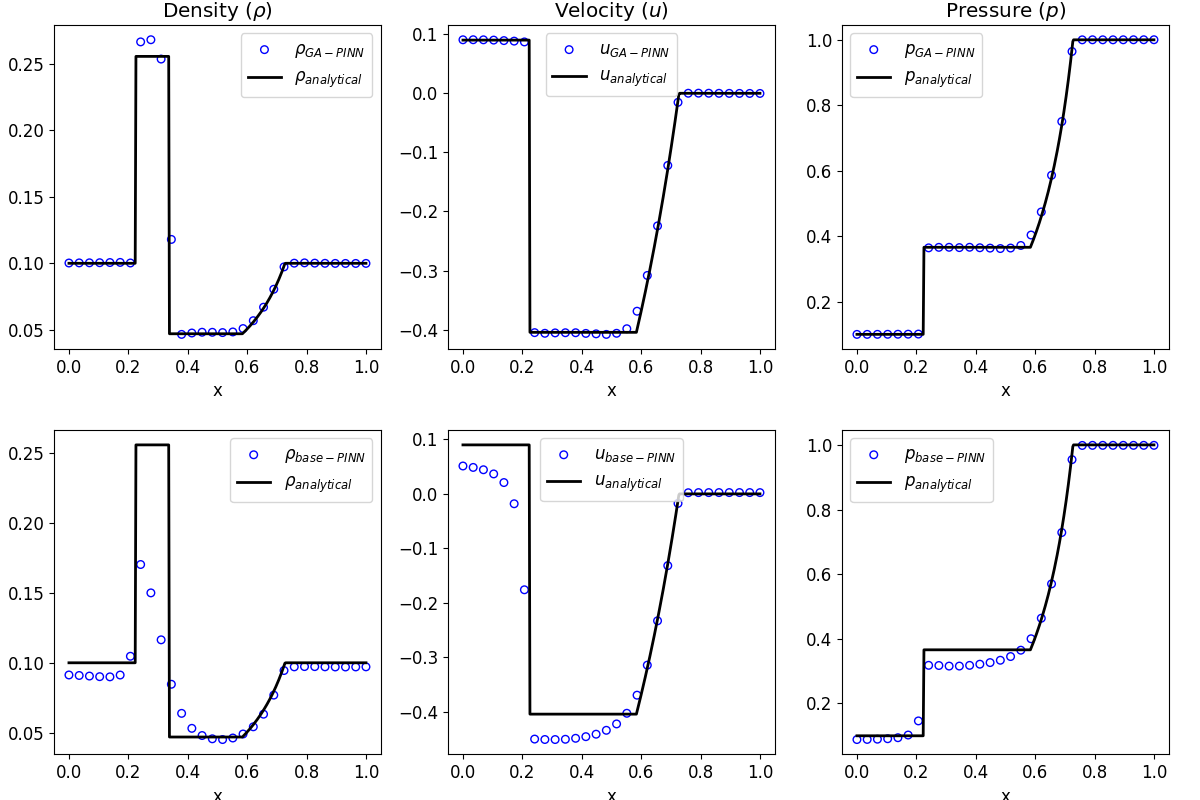} 
\caption{
Solution for the physical variables at $t=0.4$ for Problem 2. The analytical solution is represented by a solid black curve. The solutions obtained with our GA-PINNs method are shown in the top row while those for the base PINN model are displayed in the bottom row. In the two cases the solutions are represented by blue circles.}
\label{Fig_Results_Problem_2}
\end{figure}

The solution for this problem at $t=0.4$ is displayed in Figure \ref{Fig_Results_Problem_2}.
The evolution of the Riemann problem leads to a left-moving shock wave and a right-moving rarefaction wave, with a contact discontinuity in between, visible in the density and also moving to the left. The shock wave is located at $x=0.225$ and the contact discontinuity at $x=0.336$. The solution obtained with our method is shown in the top row of the figure while the prediction of a baseline PINN model is depicted in the bottom row. Again, both models have the same training properties for the common parameters reported in Table \ref{Table_Global_Results_1}, except for those strictly defined in the proposed approach.



Figure \ref{Fig_Results_Problem_2} clearly shows that the solution obtained with our GA-PINN method is significantly better than that of the base PINN model. However, there are features which indicate that the solution for the GA-PINN method is still not perfect, particularly the value of the density in the constant state between the shock wave and the contact discontinuity. To increase the performance of the neural model, the activation function \textit{softplus}\footnote{The \textit{softplus} is a smooth approximation to the well-known \textit{ReLU} function which helps the gradient to propagate backwards better because its derivative is a \textit{sigmoid} function and is therefore close to 1 as the input increases. This is a mathematically appropriate property that helps ensure that the gradient in the successive derivatives through the neural structure does not vanish and may be desirable in some cases.} has been introduced for the density and pressure outputs to better describe the contact discontinuity by achieving a less smooth behaviour. As a consequence, the initial conditions used for this particular problem  can be regarded as the most complicated to solve for a neural network of all the tests proposed in this article. This is reflected in the errors reported in Table \ref{Table_Global_Results_2}, where  the overall relative error of our methodology is $l^{2}_{\text{GA-PINN}}=0.293$, compared to an error of $l^{2}_{\text{HRSC}}=0.122$ for the HRSC model of~\cite{ref1}. However, the improvement of our methodology with respect to the PINN base model is substantial, the latter attaining an associated error of $l^{2}_{\text{base-PINN}}=0.691$. This can also be quite clearly seen in Figure \ref{Fig_Results_Problem_2} where, especially in the density variable, the effect of including a weight that suppresses the network in the discontinuities has a direct benefit for the reconstruction of the shock wave and the contact discontinuity.

\begin{figure}[t]
\centering
\includegraphics[scale=0.45]{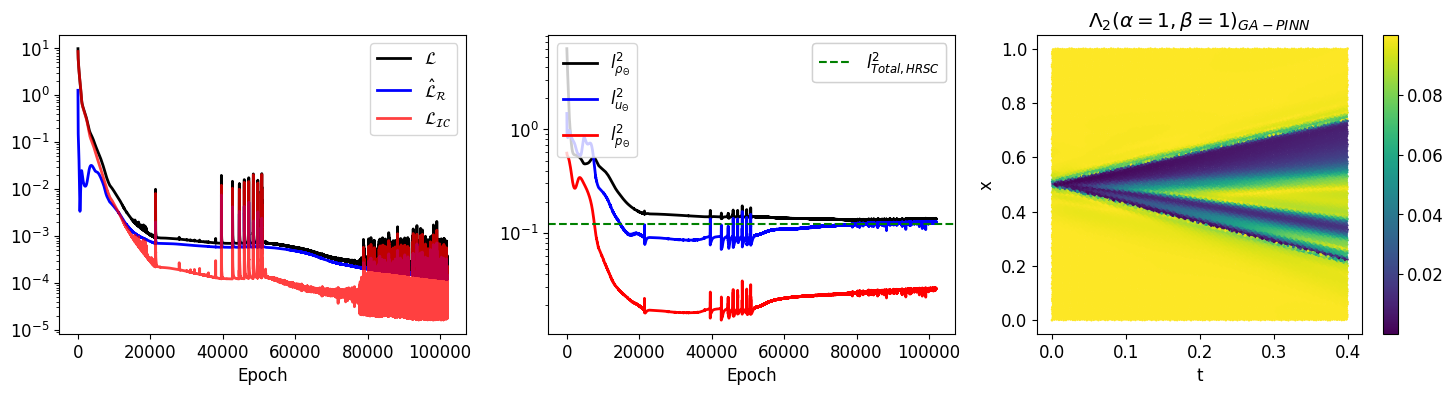} 
\caption{Evolution of the physical losses (left plot) and relative $l^{2}$ errors (middle plot) with the epoch number for Problem 2.  The analytical solution is used as ``ground truth'' to compute the relative errors. The dashed green line in the middle plot represents the corresponding error for the numerical solution obtained with the second-order central HRSC scheme of~\cite{ref1} with a mesh of 400 zones ($\Delta x=0.0025$).
The heat map on the right plot shows the $\Lambda$ function after training.}
\label{Fig_Results_Problem_2_extra}
\end{figure}
\begin{figure}[t]
\centering
\includegraphics[scale=0.4]{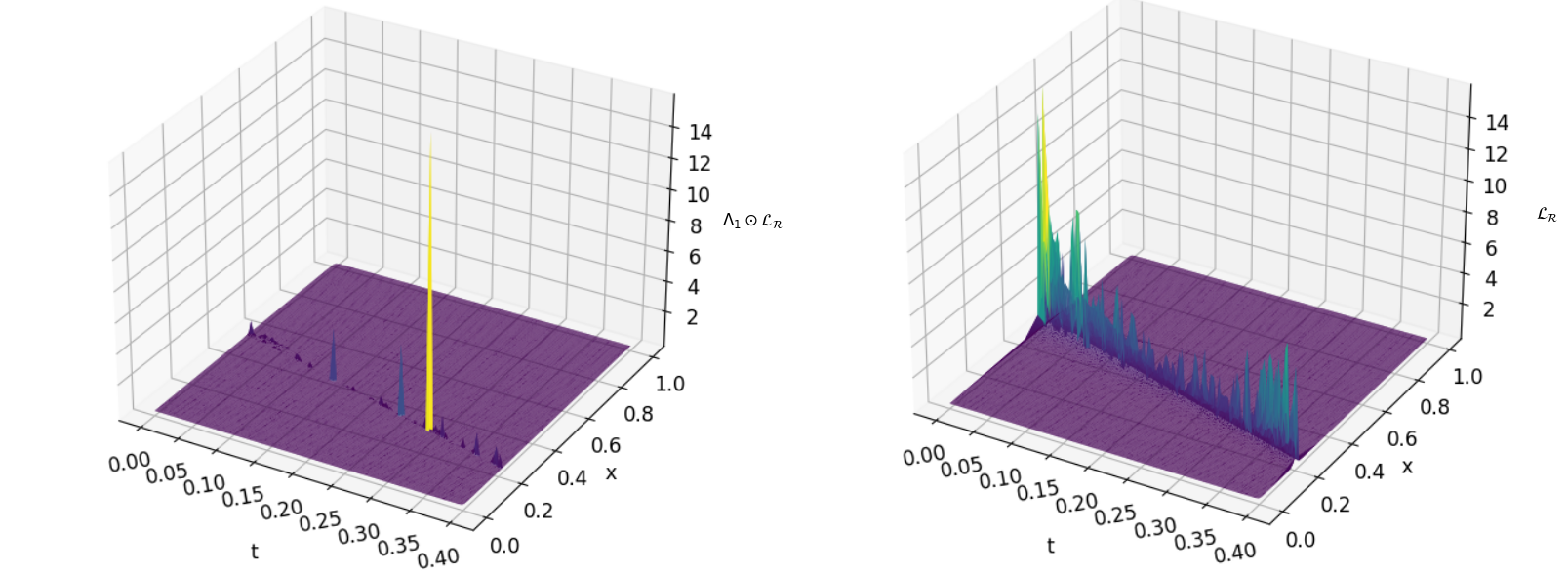} 
\caption{Three-dimensional representation of the physical losses used for the neural models in Problem 2. The left panel shows the losses when training the GA-PINN, $\Lambda_{1}\odot\mathcal{L}_{\mathcal{R}}$, while the right plot display the losses when training a baseline PINN model. The physical errors are significantly lower for our method, especially at times close to the initial time.}
\label{3D_plots_Problem_2}
\end{figure}

The left plot of Figure \ref{Fig_Results_Problem_2_extra} shows the evolution of the physical loss during training, and the middle plot displays the corresponding evolution of the relative $l^{2}$ error, for each variable. For the pressure the latter is considerably below the error for the HRSC model of \cite{ref1} (dashed green line), while the errors for the density and the velocity are similar in the two approaches. 
Furthermore, the right plot of this figure shows the function $\Lambda_{2}$ for hyperparameters $\alpha=\beta=1$. In the physical domain $(t,x)$, discontinuities can be identified as regions where $\Lambda_{2}$ exhibits relatively small values. For instance, a distinct and narrow discontinuity corresponding to the left-moving shock wave, is clearly visible, terminating at $x=0.225$ and $t=0.4$. Immediately above it, a slightly broader area is noticeable, which would correspond to a region of nearly constant values for velocity and pressure but would correspond to a contact discontinuity for the density. Finally, for large $x$ values, a considerably wider region of fluid expansion is detected.

Figure \ref{3D_plots_Problem_2} shows $\hat{\mathcal{L}}_{\mathcal{R}}(t,x)$ and $\mathcal{L}_{\mathcal{R}}(t,x)$ for our methodology and a base PINN model.
The physical errors  are considerably low for the GA-PINN (left) especially close to the initial times. In this early part of the evolution, the corresponding errors of the base model (right) reach their maximum values. Therefore, we would not expect a correct propagation of the initial conditions throughout the temporal domain. This is  what is observed in the final solution at $t=0.4$ in Figure \ref{Fig_Results_Problem_2}. A similar behaviour is observed in the remaining part of the physical domain, except at around $t=0.27$ and $x=0.35$, where our method experiences an anomalous error. Nevertheless, this feature appears as an isolated deviation at a particular spacetime point and has no substantial impact on the overall fluid evolution. Hence, it does not warrant significant concern.

\begin{figure}[t]
\centering
\includegraphics[scale=0.55]{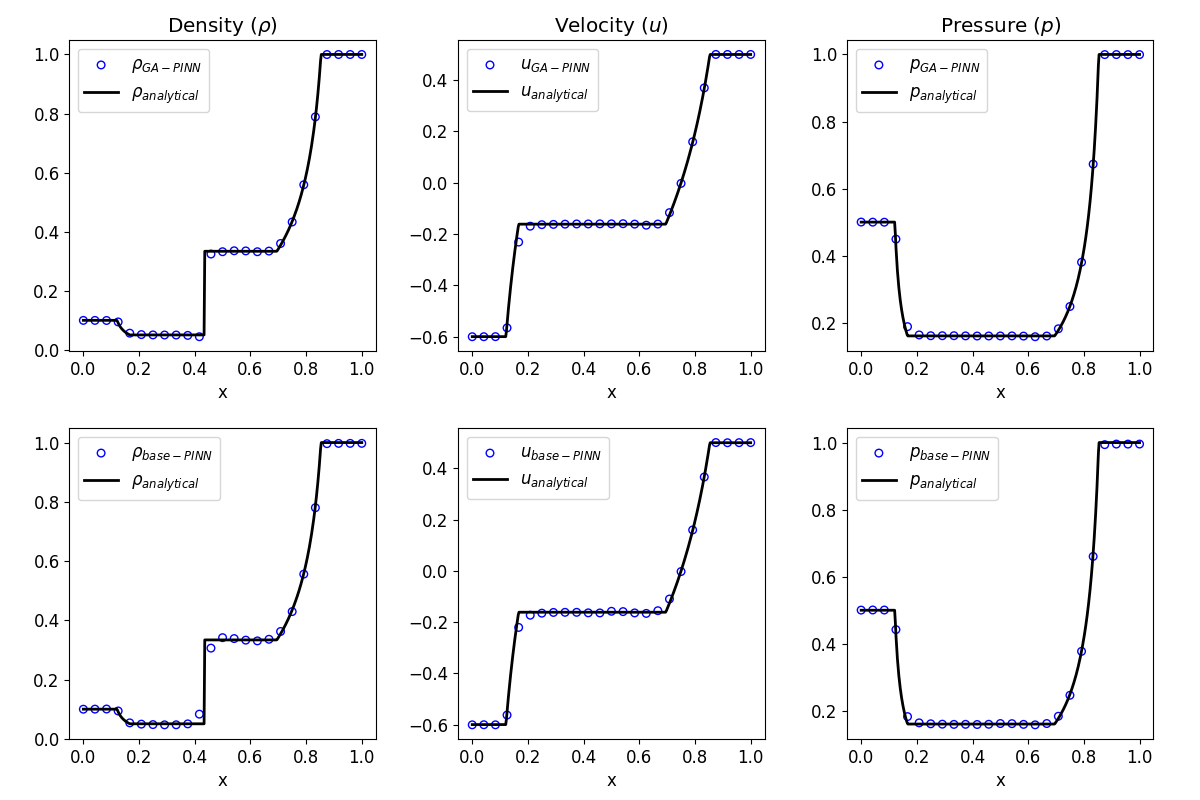} 
\caption{
Solution for the physical variables at $t=0.4$ for Problem 3. The analytical solution is represented by a solid black curve. The blue circles indicate the results obtained with our GA-PINNs method (top row) and a base PINN model (bottom row).}
\label{Fig_Results_Problem_3}
\end{figure}

\subsection{Problem 3}
\label{sec:problem3}

In the third analyzed scenario, the density and pressure exhibit a jump at $x=0.5$ while the velocity undergoes a change of sign and magnitude. This results in the fluid changing its direction of motion across the critical point $x=0.5$ at $t=0$. From left to right the (constant) value of the density increases by an order of magnitude, while the pressure doubles its value. As for problem 2, the physical domain is also $(t,x)\in[0,0.4]\times[0,1]$, although the adiabatic index of the fluid is now $\Gamma=5/3$. 
The initial conditions for this Riemann problem are summarized as:
\begin{equation}
\mathcal{U}_{0}(x)
=\left\{
\begin{array}{ll}
\rho_{0}(x)=0.1\quad\text{if $\,x\leq 0.5$,}\quad \rho_{0}(x)=1.0\quad\text{otherwise}\\
u_{0}(x)=-0.6\quad\text{if $\,x\leq 0.5$,}\quad u_{0}(x)=0.5\quad\text{otherwise}\\
p_{0}(x)=0.5\quad\text{if $\,x\leq 0.5$,}\quad p_{0}(x)=1.0\quad\text{otherwise}\\
\end{array} 
\right.
\label{ICs_problem2}
\end{equation}

Figure \ref{Fig_Results_Problem_3} shows the solution of Problem 3 at the final time $t=0.4$. In contrast to the previous two cases, in this problem there is a single discontinuity at $x=0.436$, which corresponds to the contact line of the density. Furthermore, we observe regions where the fluid expands, identified by the two rarefaction fans moving in opposite directions. The contact discontinuity is properly captured for both PINN methods, contrary to previous cases. Although our GA-PINN method still performs better than the baseline PINN model, the latter  provides an acceptable solution for this problem. This is probably due to the fact that the network only has to solve for a single discontinuity. The total relative errors computed for both approaches are similar, $l^{2}_{\text{GA-PINN}}=0.077$ and $l^{2}_{\text{base-PINN}}=0.084$, both lower than the error obtained with the second-order central HRSC scheme of~\cite{ref1}, $l^{2}_{\text{HRSC}}=0.135$.
 
\begin{figure}[t]
\centering
\includegraphics[scale=0.45]{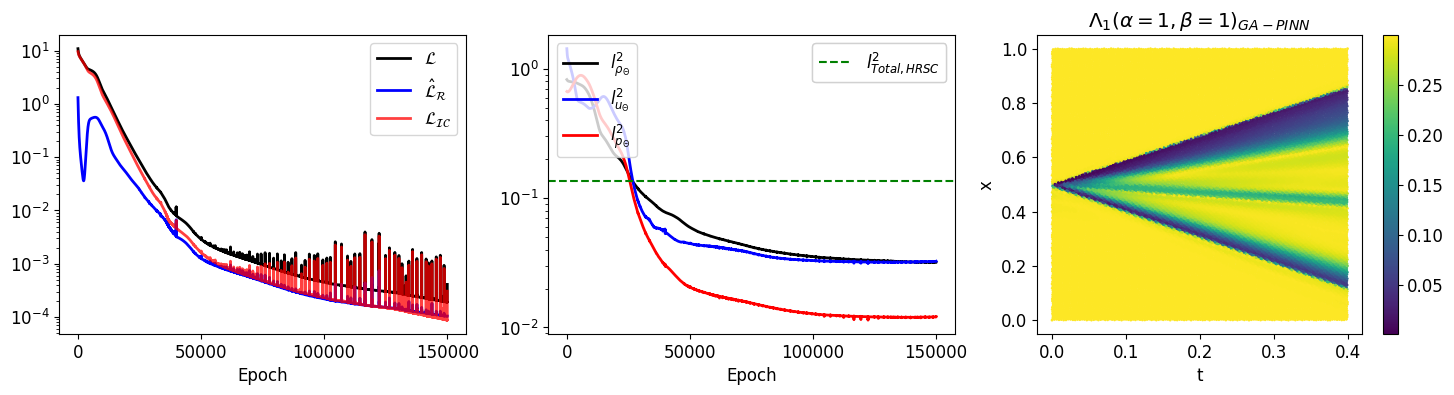} 
\caption{Evolution of the physical losses (left plot) and relative $l^{2}$ errors (middle plot) with the number of epochs for Problem 3.  The analytical solution of this Riemann problem is used as ``ground truth'' to compute the relative errors. The dashed green line in the middle plot is the $l^{2}$ error for the numerical solution obtained with the second-order HRSC scheme of~\cite{ref1} with a grid of 400 zones ($\Delta x=0.0025$). The right plot shows the $\Lambda$ function after training.}
\label{Fig_Results_Problem_3_extra}
\end{figure}

\begin{figure}[t]
\centering
\includegraphics[scale=0.4]{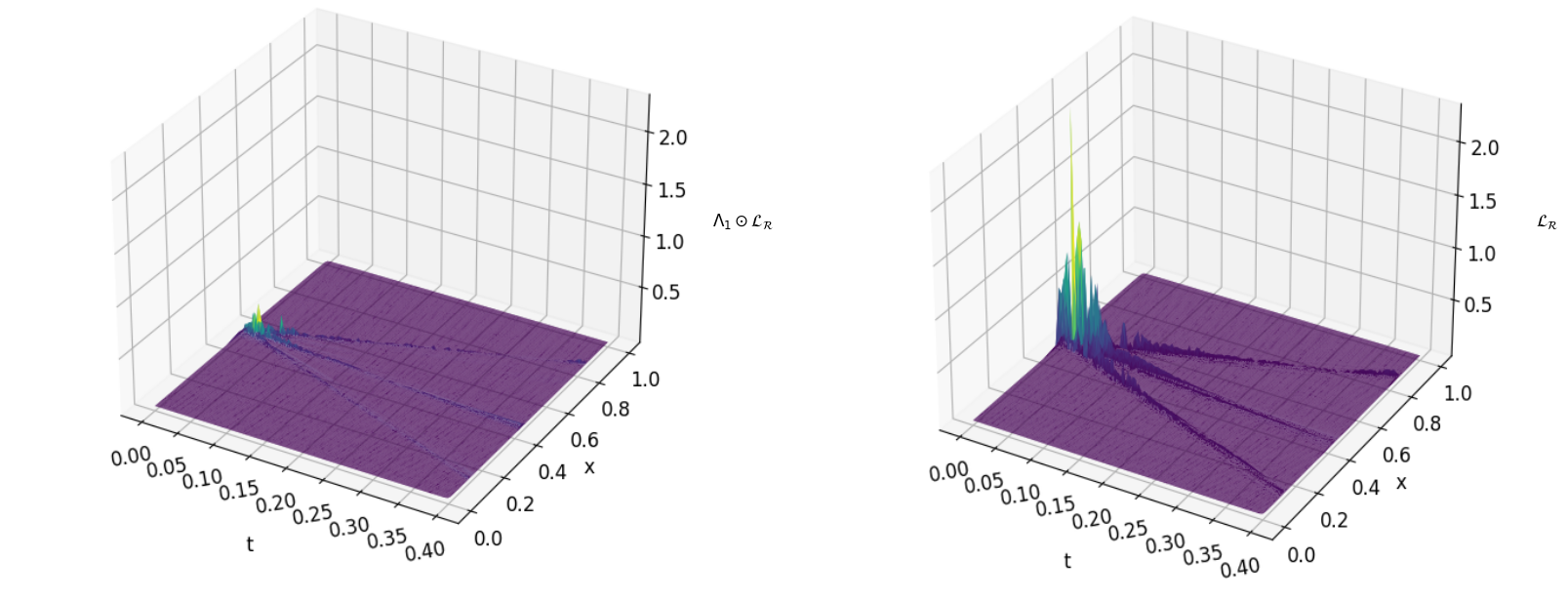} 
\caption{Three-dimensional representation of the physical losses of the neural models used in Problem 3. The left panel shows the losses when training the GA-PINN model while the right plot display the losses when training a baseline PINN model. The physical errors are significantly lower for our method, especially in the early part of the evolution.}
\label{3D_plots_Problem_3}
\end{figure}

Figure \ref{Fig_Results_Problem_3_extra} illustrates the behaviour of the physical losses ($\mathcal{L}$) and the relative $l^{2}$ errors during the training process. The physical losses (left panel) decrease monotonically in the first epochs and become ``noisy'' towards the end of the training. However, this reduction in the physical losses does not correspond to a similar reduction in the relative $l^{2}$ errors (middle panel). In particular, the physical variables have converged already after about 100,000 training iterations. The final relative errors for density and velocity are similar while the pressure is computed more accurately, as indicated by the much lower $l^{2}$ value for this variable. As mentioned before, for this problem our method produces error values that are comparable to those of a baseline PINN model. Nevertheless, this fact should not preclude the application of our methodology in less challenging problems such as problem 3, as the lack of a significant improvement does not imply a counterproductive result. The successful solution of this particular case shows the generalizable properties of our method, notwithstanding the possibility of limited benefit in certain scenarios. The right panel of Figure \ref{Fig_Results_Problem_3_extra} shows a spacetime representation of the function $\Lambda_{1}$, with hyperparameters $\alpha=\beta=1$. The contact discontinuity in the density is shown as a green line in the central part of the figure. The zones in the lowest and uppermost sections of the plot indicate the regions of continuous expansion with a steep gradient. 

Figure \ref{3D_plots_Problem_3} depicts the errors $\hat{\mathcal{L}}_{\mathcal{R}}$ and $\mathcal{L}_{\mathcal{R}}$ 
as a function of $t$ and $x$ 
for both our proposed methodology (left panel) and the baseline PINN model (right panel). The patterns observed are similar to what we found in the previous two problems. Discontinuity regions (in the variables themselves or in their first derivatives) lead to more visible errors in this figure. Those are significantly less prominent for $\hat{\mathcal{L}}_{\mathcal{R}}(t,x)$, especially close to $t=0$. This leads to a reasonable solution in the whole physical domain using our GA-PINN method.

\subsection{Problem 4: Reflection test}
\label{sec:problem4}

The last problem we study concerns the reflection of a relativistic fluid. For this test the density and pressure fields are constant over the entire spatial domain, while the velocity field undergoes a sign change with no change in magnitude. At $t=0$ the opposite-moving fluids hit against each other at $x=0.5$. This collision produces two shock waves moving toward $x<0.5$ and $x>0.5$, respectively, leaving behind a constant state of shocked, heated fluid at rest. The physical domain for this problem is $(t,x)\in[0,0.4]\times[0,1]$ and the adiabatic exponent of the equation of state is   $\Gamma=5/3$. The initial conditions read: 
\begin{equation}
\mathcal{U}_{0}(x)
=\left\{
\begin{array}{ll}
\rho_{0}(x)=1.0\quad\forall x\\
u_{0}(x)=0.5\quad\text{if $\,x\leq 0.5$,}\quad u_{0}(x)=-0.5\quad\text{otherwise}\\
p_{0}(x)=1.0\quad\forall x \, .\\ 
\end{array} 
\right.
\label{ICs_problem3}
\end{equation}

\begin{figure}[t]
\centering
\includegraphics[scale=0.50]{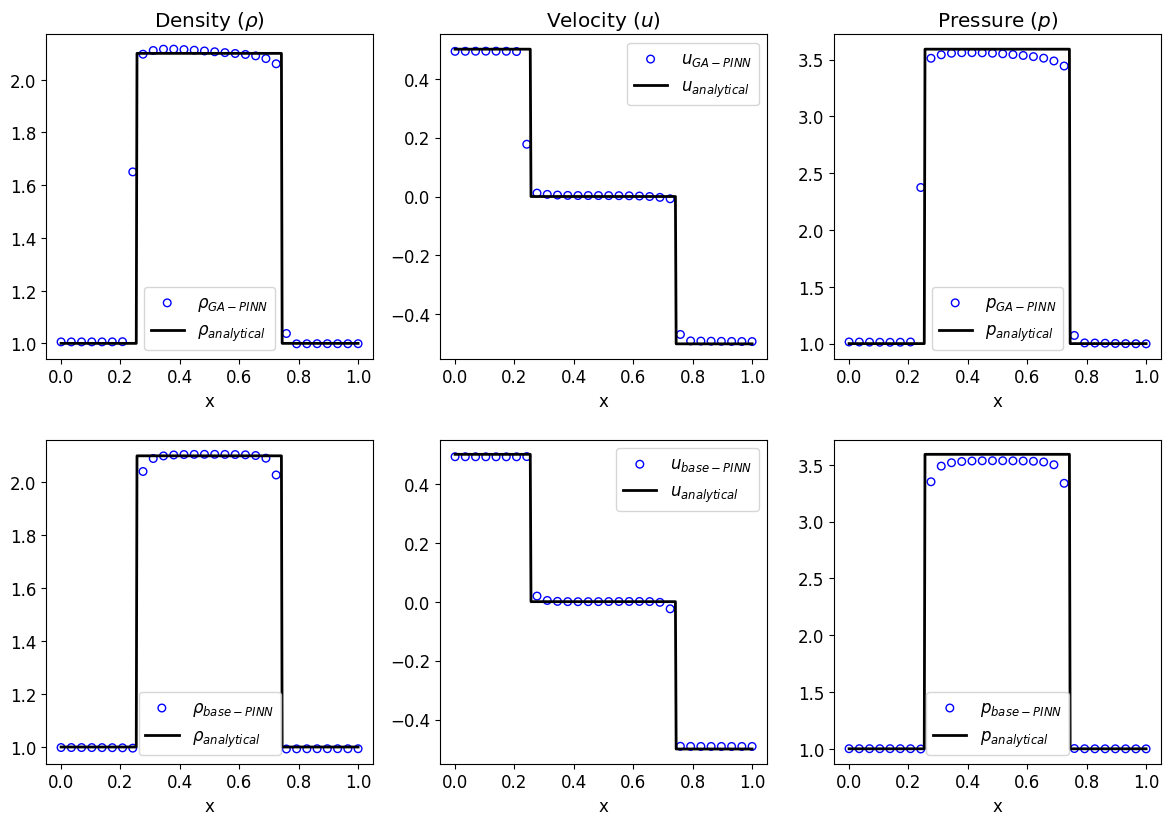} 
\caption{Solution for the physical variables at $t=0.4$ for Problem 4. The analytical solution is represented by a solid black curve and the solutions obtained with the neural networks are indicated by blue circles, those for the GA-PINN method  shown in the top row and those for the base PINN model  displayed in the bottom row.}
\label{Fig_Results_Problem_4}
\end{figure}

The solution for this problem at $t=0.4$ is shown in Figure \ref{Fig_Results_Problem_4}. The two shock waves are located at $x=0.253$ and $x=0.744$, and are visible in all fluid variables. To solve this problem it is necessary to use an unconstrained activation function, such as \textit{ReLU} or \textit{softplus}, especially for the density and the pressure. Both neural models capture properly the propagation of the shock waves and the associated jumps. However, a notable difference between our method and a base model is the latter's limited ability to accurately resolve the corners around discontinuities. This deficiency can lead to an accumulation of errors in those regions, which can degrade the overall performance score, even if the discontinuities themselves are adequately captured. Indeed, the computed overall relative errors are $l^{2}_{\text{GA-PINN }}=0.108$ and $l^{2}_{\text{base-PINN}}=0.440$, which shows once again that our proposed methodology is more accurate than the base PINN model. Moreover, the GA-PINN model also has a lower relative $l^{2}$ error than the HRSC scheme of~\cite{ref1} which yields a value of $l^{2}_{\text{HRSC}}=0.145$.  

\begin{figure}[t]
\centering
\includegraphics[scale=0.45]{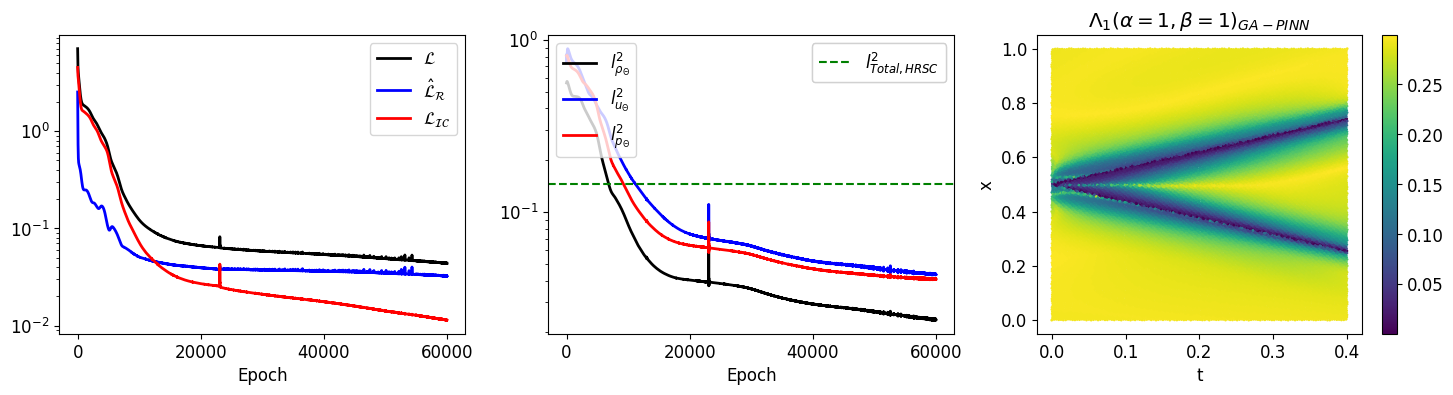} 
\caption{Evolution of the physical losses (left panel),  $\hat{\mathcal{L}}_{\mathcal{R}}$, $\mathcal{L}_{\mathcal{ IC }}$, and their sum $\mathcal{L}$, and of the relative $l^{2}$ errors (middle panel) with the number of epochs used to train the neural networks for Problem 4. The $l^{2}$ error of the HRSC scheme of \cite{ref1} with  a grid of 400 points ($\Delta x=0.0025$) is shown  with a dashed green line in the middle plot. The panel on the right shows the spacetime evolution of the $\Lambda$ function.}
\label{Fig_Results_Problem_4_extra}
\end{figure}

\begin{figure}[t]
\centering
\includegraphics[scale=0.4]{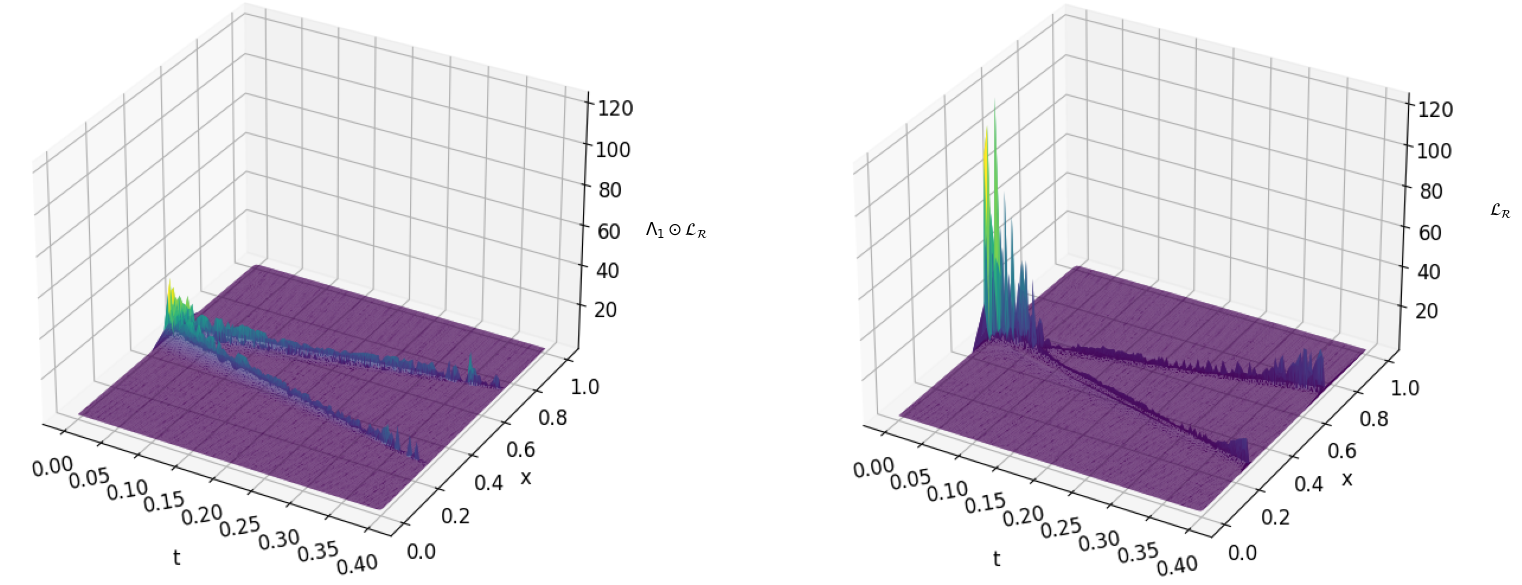} 
\caption{
Three-dimensional representation of the physical losses in Problem 4. The left panel shows the losses when training the GA-PINN, $\Lambda_{1}\odot\mathcal{L}_{\mathcal{R}}$, while the right plot display the corresponding losses for a baseline PINN model.}
\label{3D_plots_Problem_4}
\end{figure}

As done for the previous problems, we show in Figure \ref{Fig_Results_Problem_4_extra} the dependence of the physical losses and of the $l^{2}$ errors with the number of epochs. The behaviour of these quantities is as expected and, in particular, the values attained for the $l^{2}$ errors (middle plot) are well below those obtained with a second-order HRSC scheme with 400 zones. The heat map shown in the right panel of this figure, which represents the function $\Lambda_{1}$ for hyperparameters $\alpha=\beta=1$, depicts the spacetime location of the two discontinuities present in solution. It can be observed that the discontinuities evolve symmetrically away from the point $x=0.5$, which is the location where the two halves of the fluid initially collide. This is is also evident in the three-dimensional plots of the physical loss functions depicted in Figure \ref{3D_plots_Problem_4}. Similar to what we find for the previous problems, our GA-PINN methodology minimizes the errors in the initial time and in the early stages of the evolution, which leads to a generally better performance for the entire physical spacetime domain. 



\section{Discussion and conclusions}
\label{sec:discuss}

In this paper we have presented a new methodology based on Physics-Informed Neural Networks (PINNs), called Gradient-Annihilated PINNs (GA-PINNs), for solving systems of partial differential equations admitting discontinuous solutions. The performance of our GA-PINN model has been demonstrated by solving Riemann problems in special relativistic hydrodynamics, a particularly challenging physical system. To the best of our knowledge, this is the first time PINNs are employed to solve the system of equations of relativistic hydrodynamics. Therefore, the work reported in this paper has  extended to the relativistic regime earlier studies with PINNs in the context of the classical Euler equations. 

The proposed method for solving PDEs with discontinuities has a simple but intriguing motivation. It establishes a link between the challenge neural networks face when dealing with regions of steep gradients and the improbability of directly extending the universal approximation theorem \citep{ref26,ref27,ref28} without modifying the neural model to improve its performance at discontinuities. Although PINNs have produced remarkable results in solving a range of physical problems, researchers agree that this approach often over-smooths physical variables, resulting in a limited ability to account for discontinuities. It is nevertheless possible to make many and varied changes to a PINN model to adapt it to our needs.

In particular, our GA-PINN methodology focuses on two key aspects. Firstly, it is assumed that the activation functions utilized in both the internal layers and the output physical layer of the neural network exhibit dynamism whereby the set of slopes, denoted as $\{a_{k}\}_{ k=1}^{K}$, is amenable to training and subsequent minimization by the network. Additionally, it is crucial to assess and determine the most suitable activation function, or set of functions, for a given problem, as not all activation functions are physically tenable. For example, activation functions that consider velocities greater than or equal to the speed of light, will encounter difficulties from a relativistic standpoint. Secondly, we consider a modification of the residual loss function presented in equation (\ref{eq13}), which is a crucial component of the general physical loss employed for training neural models via \textit{soft enforcement}. The proposed modification involves introducing a weight function denoted as $\Lambda$, which primarily depends on the gradients of the physical variables of interest and will be inversely proportional to them, such as the ones defined in (\ref{eq15}) and (\ref{eq18}). This weight function enables the network to reduce its attention to regions with significant gradients, especially those with discontinuities. Our hypothesis is that the network enhances its performance in resolving these regions by selectively limiting learning in these areas. The numerical results have been presented as relative $l^{2}$ errors in Figure \ref{General_Results_L2_vs_Alpha} of Section \ref{sec:results}; it is evident that the network tends to improve its performance as gradients are given more consideration.

Overall, our results have shown that our GA-PINN model correctly describes the propagation speeds of the discontinuities that appear when solving the special relativistic hydrodynamics equations and sharply captures the associated jumps. In all Riemann problems investigated, the accuracy reached by our GA-PINN model has proven comparable to that obtained with the central, second-order, shock-capturing scheme of~\cite{ref1} and it is consistently higher than the accuracy achieved by a baseline, out-of-the-box PINN algorithm. Moreover, our approach  entirely sidesteps one of the main drawbacks of grid-based solutions of the relativistic hydrodynamics equations, the so-called ``primitive variable recovery''. In the case of relativistic hydrodynamics, either special or general, the determination of the primitive variables from the conserved variables (i.e.~the state vector of the PDE system) is usually not in closed form and must be performed through a root-finding procedure~\cite{Font:1994,FontLRR}. The success of such an approach is not always guaranteed, particularly for highly relativistic flows or when dealing with microphysical (tabulated) equations of state describing neutron star matter. The possibility of avoiding altogether the necessity of having to implement the recovery of the primitive variables  highlights an important benefit of our GA-PINN model. The reason is because the neural networks are capable of working with both sets of variables simultaneously in the internal calculations they perform.

One situation that remains to be fully exploited with our GA-PINN method involves ultra-relativistic flows, i.e.~fluids with $W\sim 10$ and higher. HRSC schemes written in conservation form have long proven to be well suited in dealing with such ultra-relativistic flows (see e.g.~\cite{ref39,FontLRR} and references therein). Such an extreme situation, however, leads to greater complexity in learning through a neural model. Minor deficiencies encountered when learning these ultra-high velocities could lead to significantly greater problems. To solve this subset of problems in relativistic hydrodynamics, it might be advisable to modify the neural architecture by considering a specialized neural subnetwork for each physical variable, rather than a single one whose weights are used for solving all variables. Therefore, by having a separate set of trainable variables for each physical output, the network could better learn these extreme velocities and lead to a better solution. Another option to consider is to impose the initial conditions from the start of the training process using ``hard enforcement''. This essentially involves constructing an additional small neural network that focuses solely on learning the initial conditions from the beginning. These potential extensions of the work presented here  could help  solve situations with ultra-relativistic velocities using deep learning. These are extensions that we plan to investigate in our future research.

The performance of neural networks when handling discontinuities seems to benefit from various forms of suppression of learning, as our experiments have shown. However, the experimental nature of our results and their variability depending on the initial conditions and choice of hyperparameters, clearly indicate that further improvements of our methodology are still possible. Nevertheless, the method reported in this paper already provides a solid foundation for future work in this area and constitutes a valuable tool to model relativistic flows in fields of physics  characterized by the prevalence of discontinuous solutions, such as astrophysics and particle physics.

\section*{Acknowledgements}

We thank Jos\'e Mar\'{\i}a Mart\'{\i} for kindly providing the code to compute the exact solution of the Riemann problem for special relativistic hydrodynamics.
The authors gratefully acknowledge the computer resources at Artemisa, funded by the European Union ERDF and Comunitat Valenciana as well as the technical support provided by the Instituto de Fisica Corpuscular, IFIC (CSIC-UV).
R. RdA is supported by PID2020-113644GB-I00 from the Spanish Ministerio de Ciencia e Innovación.
ATF and JAF are supported by the Spanish Agencia Estatal de Investigación (grant PID2021-125485NB-C21 funded by MCIN/AEI/10.13039/501100011033 and ERDF A way of making Europe).  AFS and JDMG are partially supported by the agreement funded by the European Union, between the Valencian Ministry of Innovation, Universities, Science and Digital Society, and the network of research centers in Artificial Intelligence (Valencian Foundation valgrAI). It has also been funded by the Valencian Government grant with reference number CIAICO/2021/184; the Spanish Ministry of Economic Affairs and Digital Transformation through the QUANTUM ENIA project call – Quantum Spain project, and the European Union through the Recovery, Transformation and Resilience Plan – NextGenerationEU within the framework of the Digital Spain 2025 Agenda.

\bibliographystyle{unsrt}
\bibliography{references}

\end{document}